\documentclass[pre,aps,10pt,preprintnumbers,twocolumn,amsmath,amssymb,floatfix,superscriptaddress,nofootinbib]{revtex4-1} 

\usepackage{graphicx}
\usepackage{dcolumn}
\usepackage{amsmath,mathrsfs,amssymb,amsthm}  
\usepackage{hhline} 
\usepackage{wasysym,enumerate,color}
\usepackage{epstopdf} 
\usepackage{bm}% bold math   
\usepackage{color}
\usepackage[colorlinks=true,citecolor=blue,linkcolor=blue,urlcolor=blue]{hyperref}
\definecolor{darkgreen}{rgb}{0,0.5,0}
\definecolor{orange}{rgb}{1,0.5,0}
\definecolor{grey}{rgb}{.6,.6,.6}

\newcommand{\be}{\begin{equation}}
\newcommand{\ee}{\end{equation}}
\newcommand{\beq}{\begin{eqnarray}}
\newcommand{\eeq}{\end{eqnarray}}

\usepackage[caption=false]{subfig}

\usepackage{epstopdf}
\usepackage{hyperref}

\definecolor{darkgreen}{rgb}{0,0.5,0}
\definecolor{orange}{rgb}{1,0.5,0}
\definecolor{grey}{rgb}{.6,.6,.6}

\begin{document}

\def\be{\begin{equation}}
\def\ee{\end{equation}}
\def\eqf{\eqref}

\def\bes{\begin{subequations}}
\def\esu{\end{subequations}}

\newcommand{\expct}[1]{\left\langle #1 \right\rangle}
\newcommand{\expcts}[1]{\langle #1 \rangle}
\newcommand{\ud}          {\mathrm d}
\newcommand\eps           {\varepsilon}
\newcommand\w           {\omega}
\newcommand\fii           {\varphi}
\newcommand\mc            {\mathcal}
\newcommand\LL            {Lieb--Liniger }
\newcommand\p             {\partial}
\newcommand\lam		{\lambda}
\newcommand\psid          {\psi^{\dagger}}
\renewcommand\th          {\theta}
\newcommand\kb            {k_\text{B}}
\newcommand \rhop       {\rho^{\text{(r)}}}
\renewcommand{\vec}[1]   {|#1\rangle}
\newcommand{\cev}[1]   {\langle#1|}

\newcommand{\fR}   {f_{\text{R}}}
\newcommand{\fL}   {f_{\text{L}}}
\newcommand{\TR}   {T_{\text{R}}}
\newcommand{\TL}   {T_{\text{L}}}
\newcommand{\rhoR}   {\rho_{\text{R}}}
\newcommand{\rhoL}   {\rho_{\text{L}}}
\newcommand{\qR}   {q_{\text{R}}}
\newcommand{\qRl}   {q_\text{R}^-}
\newcommand{\qRr}   {q_\text{R}^+}
\newcommand{\qLl}   {q_\text{L}^-}
\newcommand{\qLr}   {q_\text{L}^+}
\newcommand{\qL}   {q_{\text{L}}}
\newcommand{\ql}   {q^{-}}
\newcommand{\qr}   {q^{+}}
\newcommand{\NR}   {{N_{\text{R}}}}
\newcommand{\NL}   {{N_{\text{L}}}}
\newcommand{\xR}   {\bar x^{(\text{R})}}
\newcommand{\xL}   {\bar x^{(\text{L})}}
\newcommand{\tR}   {\bar t^{(\text{R})}}
\newcommand{\tL}   {\bar t^{(\text{L})}}

\newcommand{\vev}[1]{\left\langle #1 \right\rangle}

\frenchspacing

\title{Semiclassical theory of front propagation and front equilibration following an inhomogeneous quantum quench}

\author{M\'arton Kormos}
\affiliation{BME-MTA Statistical Field Theory Research Group, Institute of Physics, Budapest University of Technology and Economics,
H-1111 Budapest, Hungary}
\author{C\u at\u alin Pa\c scu Moca}
\affiliation{Department of Physics, University of Oradea, 410087, Oradea, Romania}
\affiliation{BME-MTA Exotic Quantum Phase Group, Institute of Physics, Budapest University of Technology and Economics,
H-1111 Budapest, Hungary}
\author{Gergely Zar\' and}
\affiliation{BME-MTA Exotic Quantum Phase Group, Institute of Physics, Budapest University of Technology and Economics, H-1111 Budapest, Hungary}

\date{\today}

\begin{abstract}
We use a semiclassical approach to study  out of equilibrium dynamics and transport  in quantum systems with massive quasiparticle excitations having internal quantum numbers. In the universal limit of low energy quasiparticles, the system is described in terms of a classical gas of colored hard-core 
particles. Starting from an inhomogeneous initial state, in this limit we give analytic expressions for the space and time dependent spin density and spin current profiles.
 Depending on the initial state, the spin transport is found to be ballistic or diffusive. In the ballistic case we identify a ``second front'' that moves more slowly than the maximal quasiparticle velocity. Our analytic results also capture 
the diffusive broadening of this ballistically propagating front. To go beyond the universal limit, we study the effect of non-trivial scattering processes in the $O(3)$ non-linear sigma model by performing Monte Carlo simulations, and  observe local equilibration around the second front in terms of the densities of the particle species.
\end{abstract}

\maketitle

\section{Introduction}

Out of equilibrium dynamics of quantum many-body systems have been in the focus of research 
in the last decade \cite{Cazalilla2010,Polkovnikov2011,Eisert2015,Calabrese2016,Vasseur2016}. Despite the great deal of progress and effort devoted to the subject,
 the mechanisms underlying transport phenomena in integrable and non-integrable systems are still not well understood.

A striking instance is provided by the presence of anomalous transport in integrable systems. For example, both diffusive and sub-diffusive spin transport have been observed numerically in the XXZ spin chain \cite{Gobert2005,Sirker2009,Znidaric2011,Steinigeweg2011,Karrasch2014,Ljubotina2017,Ljubotina2017a}. The appearance of diffusion in integrable systems is rather enigmatic as one would expect ballistic transport due to the ballistically propagating stable quasiparticles. 
Over the last year, a generalized hydrodynamical (GHD) approach has been developed that captures 
ballistic transport in Bethe Ansatz integrable systems \cite{Bertini2016,Castro-Alvaredo2016a,Doyon2017d,DeLuca2016,Sotiriadis2016,Ilievski2017a,Bulchandani2017,Bulchandani2017a,Doyon2017a,Doyon2017,Doyon2017c,Doyon2017e,Doyon2017b,Piroli2017,Ilievski2017,Bulchandani2017b,Collura2017,Cao2017,Bertini2017,Bastianello2017}. This approach is, however, unable to account for non-ballistic, and in particular, diffusive 
transport at its current stage of development.

Here we intend to pursue another, \emph{semiclassical} route to understand non-equlibrium steady state physics, 
an approach that has been successfully applied to compute dynamical correlation functions both at finite temperature~\cite{Sachdev1996,Sachdev1997,Damle2005,Rapp2006,Rapp2008} and out of equilibrium after a quantum quench \cite{Rieger2011,Evangelisti2013,Kormos2015,Moca2016}. 
This approach is  applicable to gapped one dimensional systems with quasiparticles possessing some topological or 
symmetry-protected internal quantum numbers $\mu$ which we shall refer to in what follows as `spin'.  
The meaning and possible values of $\mu$ differ from model to model: in quantum rotor models, for example,  $\mu$ corresponds to the angular momentum 
$l_z=-l,\dots,l$ of the quasiparticles~\cite{Sachdevbook}, in the sine--Gordon model it refers to the topological charge $\tau=\pm$~\cite{Damle2005,Kormos2015}, in the quantum Potts model 
it labels domain walls or residual  permutation symmetry~\cite{Rapp2006}, while in the spin-1 Heisenberg model or other similar spin models~\cite{Solyom1987} 
and non-linear sigma models~\cite{Sachdev1997,Evangelisti2013} it describes the spin of the  quasiparticles.  

\begin{figure}
\includegraphics[width=0.45\textwidth]{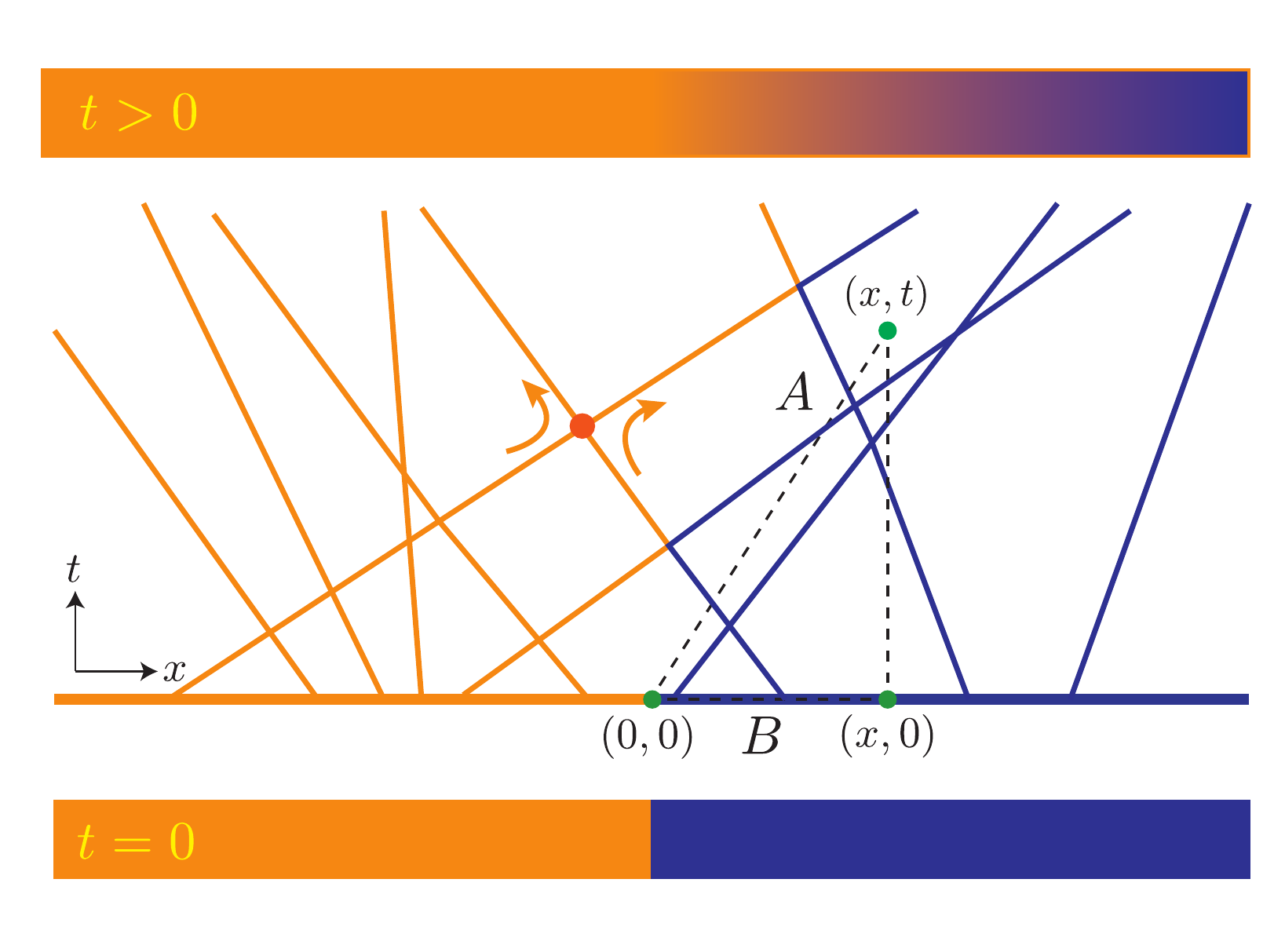}
\caption{\label{sketch} Sketch of a semiclassical configuration given by the space-time trajectories of particles.  Arrows indicate the actual trajectory of the reflected particle. Hotter particles typically create a front moving ballistically from left to right in the figure.}
\end{figure}

In this work we apply  semiclassical and hybrid semiclassical approaches to investigate  equilibration and the formation of  non-equilibrium steady 
states (NESS). The physical setup we study is the so-called partitioning protocol or tensor product initial state displayed in Fig.~\ref{sketch}, where two-semi infinite systems with different quasiparticle velocity and `spin' distributions are suddenly joined. 
This setup has been used to study energy and spin transport in various systems \cite{Antal1999,Karevski2002,Gobert2005,Calabrese2008,Lancaster2010a,Bernard2012,Sabetta2013,Karrasch2013,DeLuca2013,DeLuca2014,Collura2014a,Doyon2014b,Eisler2014a,Vasseur2015a,Bernard2016a,Viti2015,Kormos2017,Perfetto2017,Eisler2016a,Bertini2016,Castro-Alvaredo2016a,Biella2016,Bulchandani2017,Doyon2017a,Ljubotina2017,DeLuca2017,Ilievski2017a,Collura2017,Piroli2017,Alba2017,Bertini2017a,Bertini2017,Bastianello2017}.

In the semiclassical approach stable quasiparticles follow classical trajectories, while collisions are governed by quantum mechanics.  
In  the so-called \emph{universal} (low momentum) \emph{limit} the  scattering matrix of these quasiparticles becomes fully \emph{reflective}.
In this peculiar limit quasiparticles behave  in many ways as hard-core billiard balls, and their various 
correlation functions   can be computed analytically even under non-equilibrium circumstances~\cite{Rieger2011,Evangelisti2013,Kormos2015}. 
We first focus on this universal limit and  report analytic closed form expressions for the spin density  and spin current profiles valid for arbitrary times in the thermodynamic limit. We verify these analytical results by detailed Monte Carlo simulations.
 
Naively one would expect the semiclassical approach to  be able to describe only ballistic behavior. However, this is not the case if 
one is interested in the transport of internal degrees of freedom. Indeed, we identify \emph{both diffusive} and \emph{ballistic} 
spin transport in our simple model. In particular, we show that particle and `spin' densities display 
generically  ballistic `shock wave' propagation, but the front itself shows diffusive behavior.   
These results parallel the very recent results of Ref.~\cite{Medenjak2017a}, where 
the authors demonstrated  ballistic as well as diffusive behavior
in a  classical cellular automaton toy model involving charged hard-core particles and neutral 
non-interacting particles. The behavior we observe  is similar to that observed in classical hard-core models~\cite{Jepsen1965,Levitt1973,Balakrishnan2002}.  

Unfortunately, the analytical approach described in the previous paragraphs has its limitations, since
for quasiparticles of finite kinetic energy the scattering matrix is not fully reflective. Nevertheless, as demonstrated recently~\cite{Moca2016}, it  is  possible to go beyond this universal 
limit by means of a hybrid \emph{semi-semiclassical} Monte Carlo approach, 
and thereby account for the generically weak transmissive scattering events and  simulate 
the actual physical systems accurately at long times.

Our hybrid semiclassical simulations show that the generic features survive beyond the universal limit, i.e., if we allow transmissive and `spin' changing scattering processes encoded by a non-trivial scattering matrix. As an example, we analyze the $O(3)$ non-linear sigma model that provides the low energy effective description of the spin-1 Heisenberg spin chains in the Haldane gapped phase. We find that the non-trivial S-matrix opens new quasiparticle `spin' relaxation channels and leads to certain new phenomena, including the 
equilibration of quasiparticle species at the front.

The paper is organized in the following way. After specifying our model and the physical setup, we summarize the derivation of the analytical results obtained in the universal limit in Sec. \ref{sec:deriv}. The details of this calculation as well as an alternative derivation are presented in the Appendix. In Sec. \ref{sec:asym} we analyze the asymptotic behavior of the space-time profiles and identify ballistic and diffusive contributions to the spin transport. We also compare the analytical results of the universal semiclassical approach with numerical simulations.
In Sec.~\ref{sec:level1} we extend the latter to semi-semiclassical simulations that account for transmissive as well as more general processes in the $O(3)$ non-linear sigma model.
Finally, we give our conclusions and discuss possible extensions of our work in Sec. \ref{sec:concl}. 

\section{Analytic magnetization and current profiles in the universal limit}
\label{sec:deriv}

\subsection{The setup}

The out of equilibrium evolution and transport is induced by joining two semi-infinite systems that are in different 
homogeneous states, pure or mixed, which are represented in the semiclassical picture by a distribution of quasiparticle excitations. We assume that quasiparticles have the same mass on both sides, but their velocity as well as their internal `spin'  $\mu$ may have a different distribution on either side.

 Notice that the quantum number $\mu$ may, in general, correspond to any internal quantum number such as 
 charge, topological charge, SU(2) spin label,  or virtually any other internal degree of freedom.
 In the $O(3)$ non-linear sigma model, e.g.,  elementary excitations are triplet particles with spin components $S^z\equiv\mu=\pm1,0$, and  have relativistic dispersion relations $\eps(p) = \sqrt{\Delta^2+p^2c^2}.$ In this  integrable 
model, the two-body S-matrix is exactly known (see Appendix \ref{app:Smatrix}), and  in the universal limit of low momenta it becomes perfectly reflective. 
In this and the following  section we focus entirely on this \emph{universal} purely reflective limit, and derive general analytical semiclassical  results for the time evolution of the steady state.  
(Effects emerging beyond this universal limit shall be discussed in Sec.~\ref{sec:level1}).

The initial distribution of quasiparticles is characterized by the distribution functions $f_\text{L/R}(\mu,p),$ where $p$ is the momentum and the superscripts $L$ or $R$ refer to the left and right systems, respectively. Our only assumption is that the distribution function factorizes,
\be
\label{fmup}
f_\text{L/R}(\mu,p)=g_\text{L/R}(\mu)\,f_\text{L/R}(p)
\ee
with $\sum_\mu g_\text{L/R}(\mu) =1.$ This condition can be released in numerical simulations, however, it is important for the analytical solution presented in this section. 
 An example for such a factorization is given by thermal equilibrium in the presence of some external field  $h$, when 
 \be
 \label{fmuptherm}
 f_{\alpha}(\mu,p) = e^{\beta_{\alpha} h_{\alpha} \mu} e^{-\beta_{\alpha}\eps(p)} 
 \ee
where $\alpha= \text\{L,R\}.$  A semiclassical configuration is thus given by the initial locations, momenta, and spins of all  quasiparticles. 
We assume that particles are initially evenly distributed in space on each side
and their momenta and spins are drawn from the distribution $ f_\text{L/R}(\mu,p).$ 
We are interested in averages over these configurations, i.e. over the $\{x,p,\mu\}$ initial
coordinates of all  particles. Pictorially, such a configuration is represented by a set of semi-infinite straight lines 
in the $(x,t)$ plane 
(see Fig. \ref{sketch}), and by energy and momentum conservation, particles must move along segments of these lines. 
Each line starts at a random point of the $t=0$ horizontal line with a slope 
\be
v_p=\frac{\ud \eps(p)}{\ud p}
\ee
corresponding to the initial velocity of the particle and each segment
carries a label $\mu$ according to the spin of the particle.
We connect the two half systems at $t=0$ and $x=0$, and to generate transport, the momenta and the spins are drawn at $t=0$ 
from different distributions for $x>0$ and $x<0.$ 
The initial particle number and magnetization densities on the two  sides are given by
\begin{align}
n_\text{L/R} &= \sum_\mu \int\frac{\ud p}{2\pi}  f_\text{L/R}(\mu,p)=\int\frac{\ud p}{2\pi}  f_\text{L/R}(p)\,,\label{eq:n_init}\\
m_\text{L/R} &= \sum_\mu \int\frac{\ud p}{2\pi} \, \mu\, f_\text{L/R}(\mu,p) =n_\text{L/R}\sum_\mu \mu \,g_\text{L/R}(\mu) \,,
\end{align}
and the initial  polarization, i.e. the average magnetization of a single particle is 
expressed as
\be
\hat \mu_\text{L/R}\equiv\vev{\mu}_\text{L/R}= m_\text{L/R}/n_\text{L/R}=\sum_\mu \mu\,g_\text{L/R}(\mu)\,.
\ee
Notice that  straight lines in Fig.~\ref{sketch} are not the actual physical trajectories of the hard-core particles because their
spin degree of freedom gets reflected and follows complicated zig-zag trajectories (see Fig.~\ref{sketch}). It is this non-trivial motion of the particles that render the calculation of the spin current non-trivial.
Indeed, in contrast to the spin, expectation values of  fully transmitted quantities 
such as energy or particle density are easy to compute, because they propagate along the straight trajectories. Therefore, near position $x$ and at time $t$ only those particles  contribute to the current that come from the left and are faster than $x/t,$ and those that come from the right and are slower than $x/t$. In Fig.~\ref{sketch} each line carries momentum $p$, energy $\eps(p)$, and unit particle number. For example, the mean particle density and current are given by
\begin{align}
n(x,t) &= \int\frac{\ud p}{2\pi}\bigl[\Theta(x/t-v_p) f_\text{R}(p) + \Theta(v_p-x/t)f_\text{L}(p)\bigr]\,
,\label{rhop}
\\
\begin{split}
j(x,t) &= \int\frac{\ud p}{2\pi} \,v_p \, \Theta(x/t-v_p) f_\text{R}(p) 
\\
&\phantom{mmmmm}+ \int\frac{\ud p}{2\pi}  \,v_p \, \Theta(v_p-x/t)f_\text{L}(p) \,v_p\,.
\label{jp}
\end{split}
\end{align}
The energy density and energy current are given by analogous expressions, only the integrands contain an extra factor of $\eps(p).$ Note that these expressions depend solely  on the scaling variable 
\be
\xi=x/t\,,
\ee
i.e. on the ``ray'' in the $(x,t)$ plane. 
If there is a maximal quasiparticle velocity $c$, then  particle and energy currents are zero outside of the light cone, i.e. for $|\xi|>c.$
Keeping $x$ finite while sending $t\to\infty$ corresponds to the $\xi=0$ ray, identified as 
 the non-equilibrium steady state (NESS) developing at the center.

\subsection{Analytic space-time profiles}

We shall now derive closed, analytical expressions for the complete spatial and time dependence of the magnetization profile, $m(x,t)$ in the 
limit of fully reflective collisions.  To do that, we shall first compute the average magnetization $M(x,t)$ transferred through point $x$ until time $t$. Differentiation then yields immediately the magnetization current, $J(x,t) = \partial_t M$, and the change in the magnetization density $m(x,t)-m(x,0) = -\partial_x M$.

For a given configuration $\cal C$ and time $t,$
the magnetization  $M({\cal C})$ transported across point $x$ is the sum of the spins of the particles crossing the segment $[(x,0),(x,t)],$ each weighted by $\pm1$ depending on whether it crosses from the left or from the right. To compute $M(x,t)$  we must average this quantity over all possible initial quasiparticle configurations, 
$M(x,t)=\langle M({\cal C})\rangle_{\cal C}$. 

Consider now the triangle on the $(x,t)$ plane with vertices $(0,0),$ $(x,0),$ and $(x,t)$, shown in  Fig. \ref{sketch}. As magnetization is conserved by the dynamics, the total magnetization flowing into this triangle, including the inflow along the edge $B=[(0,0),(x,0)],$ must be zero \footnote{Or, thinking in terms of spatial domains, the initial magnetization of the interval $B$ must flow out at the left and the right boundaries while we shrink the interval to zero by moving the left boundary to the right one.}. This implies that $M({\cal C})$ can also be calculated as the sum of the spins (again with signs) encountered when moving along the $A=[(0,0),(x,t)]$ segment, $ M_A({\cal C})$, plus the sum of the spins along the $B$ interval, $ M_B({\cal C})$. The second quantity is related to the initial magnetization, but $ M_A$ is, in principle, complicated: although the spins of particles at $t=0$ are uncorrelated, they travel along zig-zag trajectories as a result of multiple collisions, and they can cross  segment $A$ multiple times. 

However, due to the perfectly reflective elastic collisions, the spatial sequence of the spins at any fixed time is unchanged under the time evolution. 
As a consequence, if the number of net crossings along $A$ is $s,$
$ M_A$ is equal to the negative sum of the first $|s|$ spins to the right (if $s<0$) or to the sum of the first $s$ spins to the left (if $s>0$) of the origin at $t=0$.

To obtain the expectation value of the transported magnetization, we have to average over all semiclassical configurations. As the spin and orbital degrees of freedom follow independent distributions, we can first average over the spins. This implies that the average left and right magnetizations can be used, 
yielding 
\bes
\label{Meqs}
\begin{align}
 \langle M_A \rangle_{\cal C} &=\vev{\Theta(s)s}
 \hat\mu_\text{L} +\vev{\Theta(-s)(-s)}(-\hat\mu_\text{R})\,,
\label{Mt}\\
 \langle M_B \rangle_{\cal C}&=(\Theta(x)\hat\mu_\text{R} n_\text{R} +\Theta(-x) \hat\mu_\text{L}
  n_\text{L}) x\,,\label{M0}
\end{align}
\esu
where $\vev{\dots}$ denotes averaging over the remaining orbital degrees of freedom:
\begin{multline}
\label{ave}
\vev{O} = 
\frac1{(n_\text{R}L)^{N_\text{R}}}\prod_{i=1}^{N_\text{R}}\int_0^L\ud y_i \int \frac{\ud p_i}{2\pi} f_\text{R}(p_i)
\\
\frac1{(n_\text{L} L)^{N_\text{L}}}\prod_{j=1}^{N_\text{L}}\int_{-L}^0\ud \bar y_j
\int \frac{\ud \bar p_j}{2\pi} f_\text{L}(\bar p_j)
\,O\,,
\end{multline}
where the physical quantity $O$ depends implicitly on the initial positions and momenta of the particles,
$\{y_i,p_i\}$ and  $\{\bar y_j,\bar p_j\}$, with the bar referring to particles on the left. Then the average transported magnetization is
\be
\label{M}
M(x,t) =  \langle M_A \rangle_{\cal C} + \langle M_B \rangle_{\cal C} \,.
\ee

To evaluate  $\vev{\Theta(s)s}$  and $\vev{\Theta(-s)s}$ 
we notice  that  $s$ is just a crossing number, and is simply given by  the number of straight lines from the 
 right ending up left of the point $x$ at time $t$ minus the number of lines from the left ending up right of $(x,t),$ 
\be
\label{s}
s=\sum_{j=1}^{N_\text{L}}\Theta(\bar y_j+v(\bar p_j)t-x) - \sum_{j=1}^{N_\text{R}}\Theta(x-y_j-v(p_j)t)\,,
\ee
where it is understood that $y_j>0$ and $\bar y_j<0.$

To evaluate  $\vev{\Theta(s)s}$  we rewrite the Heaviside theta function as
$\Theta(s) = \int \frac{\ud p}{2\pi} \frac{e^{ips}}{ip+\eps}.$
Now the average $\vev{e^{ips}s}$ is a product of averages over independent variables, which we can evaluate 
analytically and reexponentiate the result in a few steps to yield (see Appendix \ref{app:deriv}) 
\be
\begin{split}
\vev{\Theta(s)s} &= 2\sqrt{Q_\text{R}Q_\text{L}}e^{-Q_\text{R}-Q_\text{L}} \times
\\&\int \frac{\ud u}{2\pi} \frac{\sin(u-i\gamma)}{u-i\eps} e^{2\sqrt{Q_\text{R}Q_\text{L}}\cos (u-i\gamma)}\,,
\end{split}
\ee
where $\tanh\gamma = (Q_\text{L}-Q_\text{R})/(Q_\text{L}+Q_\text{R})$ and
\bes
\label{Qdef}
\begin{align}
Q_\text{R}(x,t) &= \int
\frac{\ud p}{2\pi} \Theta(x/t-v_p)f_\text{R}(p)(x-v_pt)\,,\\
Q_\text{L} (x,t)&= \int
\frac{\ud p}{2\pi} \Theta(v_p-x/t)f_\text{L}(p)(v_pt-x)
\end{align}
\esu
are the expectation numbers of right/left particles crossing the segment $A$ connecting the origin 
with the point $(x,t).$
Repeating the derivation for $\vev{\Theta(-s)(-s)}$ and using Eq. \eqf{Mt}, we finally obtain for $ \vev{M_A}$
\begin{multline}
\label{RES1}
 \vev{M_A} =  
2\sqrt{Q_\text{R}Q_\text{L}}e^{-Q_\text{R}-Q_\text{L}}\\
\int \frac{\ud u}{2\pi} \sin(u-i\gamma) e^{2\sqrt{Q_\text{R}Q_\text{L}}\cos (u-i\gamma)}\left(\frac{{\hat\mu_\text{L}}}{u-i\eps}  -\frac{{\hat\mu_\text{R}}}{u+i\eps} \right)\,.
\end{multline}
An alternative but maybe less transparent derivation presented in  Appendix \ref{sec:alt} yields an equivalent but possibly  
more convenient, alternative expression, 
\begin{multline}
\label{res}
 \vev{M_A}Ê=
(Q_\text{L}-Q_\text{R})\left(\Theta[Q_\text{R}-Q_\text{L}]{\hat\mu_\text{R}}+\Theta[Q_\text{L}-Q_\text{R}]{\hat\mu_\text{L}}\right)\\
+({\hat\mu_\text{L}}-{\hat\mu_\text{R}})\sqrt{Q_\text{R}Q_\text{L}}\int_1^\infty \ud z \frac{e^{-(Q_\text{R}+Q_\text{L})z}}{z} I_1\left(2\sqrt{Q_\text{R}Q_\text{L}}\,z\right)\,,
\end{multline}
where $I_1(x)$ is the modified Bessel function of the first kind.

The spin current is given by the time derivative of the total transported magnetization at position $x$, 
\be
\label{jdef}
J(t) = \p_t M(x,t)  = \p_t \vev{M_A} \,,
\ee
since $ \vev{M_B}$ is independent of time. The magnetization (spin) density $m(x,t)$ can then 
be obtained by integrating the continuity equation
\be
\label{cont}
\p_t m(x,t)  + \p_x J(x,t)  = 0
\ee
with the initial condition given by the initial state. This yields
\begin{multline}
\label{m}
m(x,t) = - \p_x M(x,t)  + \Theta(x)n_\text{R}{\hat\mu_\text{R}}+\Theta(-x)n_\text{L}{\hat\mu_\text{L}}\\
=- \p_x    \vev{M_A} \,,
\end{multline}
where we noticed that the initial condition exactly cancels the spatial derivative of $\langle M_B \rangle_{\cal C}.$
The function $ \vev{M_A}$ depends on $x$ and $t$ through $Q_\text{L}$ and $Q_\text{R}$ only, which implies that apart from an overall sign, the expression for $m(x,t)$ and $J(x,t)$ will have the same structure.
 When differentiating Eq. \eqf{res}, it is useful to change the integration variable $z$ to $u=2\sqrt{Q_\text{R}Q_\text{L}}z$ and change it back after differentiation. 
 We thus find
\begin{widetext}
\begin{multline}
\label{jmres}
\begin{pmatrix}
J(x,t)\\m(x,t)
\end{pmatrix}
 = (\nabla Q_\text{L}-\nabla Q_\text{R})\left(\Theta[Q_\text{R}-Q_\text{L}]{\hat\mu_\text{R}}+\Theta[Q_\text{L}-Q_\text{R}]{\hat\mu_\text{L}}\right)\\
+({\hat\mu_\text{L}}-{\hat\mu_\text{R}})\frac{\nabla Q_\text{R}Q_\text{L}+Q_\text{R}\nabla Q_\text{L}}{2\sqrt{Q_\text{R}Q_\text{L}}}\left(\int_1^\infty \ud z \frac{e^{-(Q_\text{R}+Q_\text{L})z}}{z} I_1\left(2\sqrt{Q_\text{R}Q_\text{L}}\,z\right)-e^{-(Q_\text{R}+Q_\text{L})}I_1(2\sqrt{Q_\text{R}Q_\text{L}})\right)\\
-({\hat\mu_\text{L}}-{\hat\mu_\text{R}})\frac{\nabla Q_\text{R}Q_\text{L}-Q_\text{R}\nabla Q_\text{L}}{2\sqrt{Q_\text{R}Q_\text{L}}}(Q_\text{R}-Q_\text{L})
\int_1^\infty \ud z e^{-(Q_\text{R}+Q_\text{L})z} I_1\left(2\sqrt{Q_\text{R}Q_\text{L}}\,z\right)\,,
\end{multline}
\end{widetext}
where 
\be
\nabla Q_\text{R/L} (x,t) \equiv
\begin{pmatrix} \p_t Q_\text{R/L} 
\\-\p_x Q_\text{R/L} \end{pmatrix}\,.
\ee
Expression \eqf{jmres} is one of the main results of the paper that gives the space-time profile of the magnetization current and density in the thermodynamic limit for arbitrary $x$ and $t.$

\section{Large time asymptotic results}
\label{sec:asym}

Although Eq.~\eqref{jmres} has an analytical beauty, it is not very transparent. To gain some physical 
insight, let us now  analyze its physical content by extracting its behavior at large times.

For later purposes, let us introduce the \emph{rates} at which particles from the 
right/left cross segment $A$ 
\be
\Gamma_\text{R/L}(x,t)\equiv Q_\text{R/L}(x,t)/t\,.
\ee
Interestingly, these rates 
depend on $x$ and $t$ only through the ratio $\xi\equiv x/t$, with the velocity variable 
 $\xi$ specifying    ``rays'' in the $(x,t)$ plane.
 Similarly, 
the derivatives 
$\p_t Q_\text{R/L}(x,t)$ and $\p_x Q_\text{R/L}(x,t)$ also depend on $\xi$ only,
\begin{align}
\label{Qder}
\p_x Q_\text{R/L} (\xi)  
&= \pm\int\frac{\ud p}{2\pi} \Theta[(\pm(\xi-v_p)]f^{R/L}(p)\,,
\\
\p_t Q_\text{R/L}(x,t) 
& = \mp\int\frac{\ud p}{2\pi} \Theta[\pm(\xi-v_p)]\;v_p\;f^{R/L}(p) \,.
\end{align}
Notice that $\p_x Q_\text{R}-\p_x Q_\text{L}$ is nothing but the particle density $n(x,t) = n(\xi)$ in Eq. \eqf{rhop} and $\p_t Q_\text{L}-\p_t Q_\text{R}$ is the particle current 
$j(x,t) = j(\xi)$ in Eq. \eqf{jp}.

\begin{figure}[t!]
\includegraphics[width=.5\textwidth]{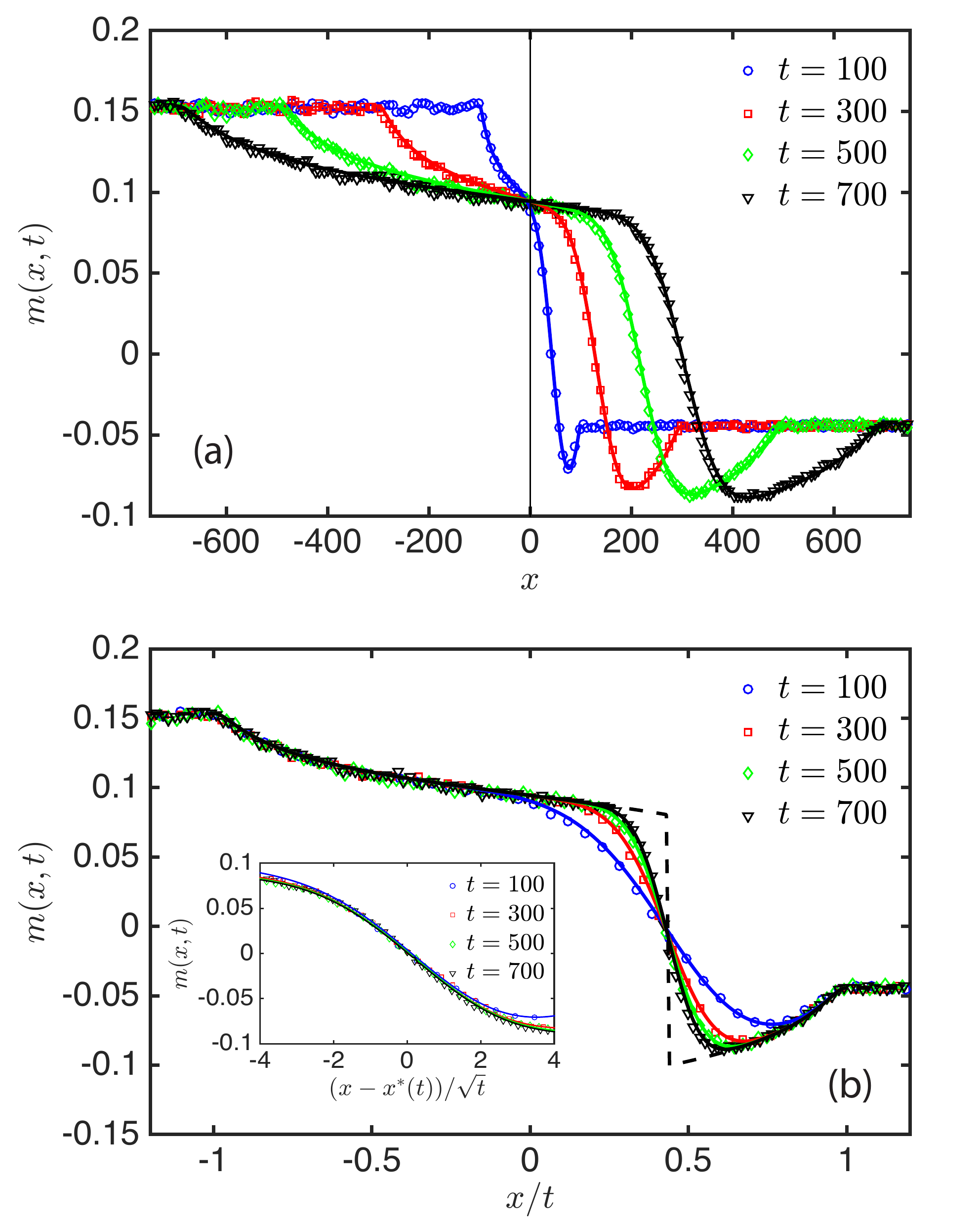}
\caption{\label{fig:mprofiles} Evolution of the magnetization profile for particles with  dispersion relation $\eps(p)=\sqrt{\Delta^2+p^2c^2}.$ The coordinate $x$ is measured in units of the Compton length $\hbar c/\Delta$ while time $t$ is measured in units of $\hbar /\Delta,$ and we set $\hbar=1$ and $c=1.$  The left/right momentum distributions are $ f_\text{L/R}(p)\sim e^{-\beta_\text{L/R}\eps(p)}$ with inverse temperatures $\Delta\;\beta_\text{L}=1,$ $\Delta\;\beta_\text{R}=2$ and average magnetizations per particle ${\hat\mu_\text{L}}=0.8,$ ${\hat\mu_\text{R}}=-1.$ 
(a) Magnetization profiles for different times  
as indicated in the legend. The analytic result \eqf{jmres} is plotted in solid lines while  symbols represent  Monte Carlo  simulations. (b)  Same data   as functions of $x/t$, demonstrating ballistic transport.  
Eq. \eqf{m_bal} is shown as a dashed line.
{\em Inset:} Magnetization profiles around the second front  as a function of $[x-x^*(t)]/\sqrt{t}$, demonstrating the diffusive broadening of the front.
}
\end{figure}

\begin{figure}[t!]
\includegraphics[width=.5\textwidth]{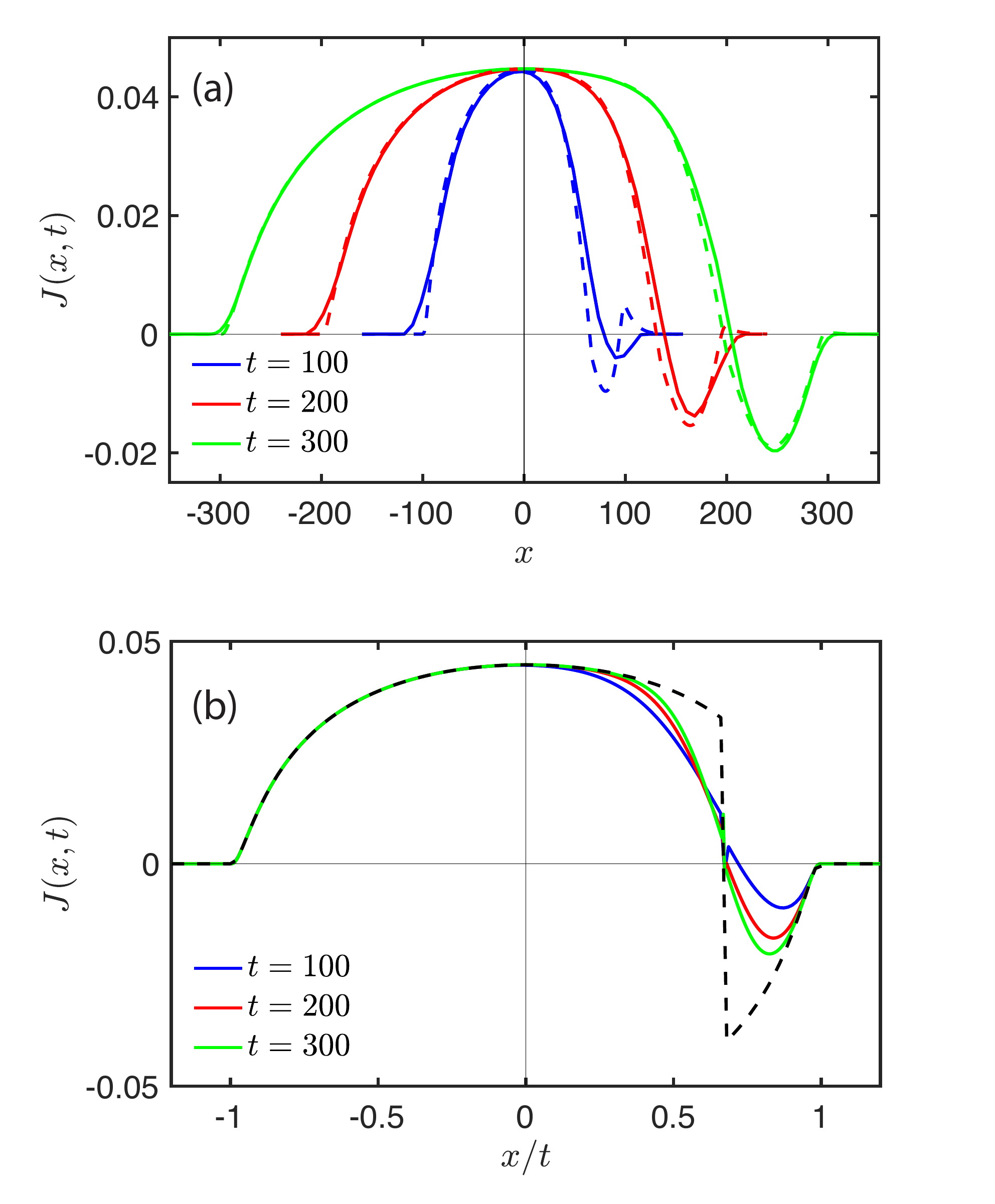}
\caption{\label{fig:jprofiles} 
Evolution of  magnetization current profile for the setup in Fig. \ref{fig:mprofiles} except that the right temperature is $\Delta\;\beta_\text{R}=3.$ (a) Magnetization current profiles for different times  
as indicated in the legend. The analytic result \eqf{jmres} is plotted in solid lines while the dashed line represent the asymptotic expression \eqf{jmas}. (b) Magnetization c plotted as function of $x/t$ to demonstrate ballistic front propagation. The ballistic result in Eq. \eqf{j_bal} is shown in dashed line.}
\end{figure}

Since we are mainly interested in the large time behavior of the magnetization density and the current profiles, and since 
 $Q_\text{R/L}$ are both proportional to $t$ for any fixed $\xi$,  
for large times we can use the asymptotic behavior of the Bessel function 
and obtain an analytical estimate   of the integrals in Eq. \eqf{jmres}  (see Appendix \ref{app:asym}) 
\begin{multline}
\label{jmas} 
\begin{pmatrix}
J(x,t)\\m(x,t)
\end{pmatrix} \\
\approx  (\nabla Q_\text{L}-\nabla Q_\text{R})\left(\Theta[Q_\text{R}-Q_\text{L}]{\hat\mu_\text{R}}+\Theta[Q_\text{L}-Q_\text{R}]{\hat\mu_\text{L}}\right)\\
+\Delta{\hat \mu}\frac{\nabla Q_\text{R}Q_\text{L}+Q_\text{R}\nabla Q_\text{L}}{2(Q_\text{R}Q_\text{L})^{3/4}}\left(\frac{e^{-R^2}}{2\sqrt\pi}
-|R|\mathrm{erfc}|R|\right)\\
-\Delta{\hat \mu}\frac{\nabla Q_\text{R}Q_\text{L}-Q_\text{R}\nabla Q_\text{L}}{4(Q_\text{R}Q_\text{L})^{3/4}}\left(\sqrt{Q_\text{R}}+\sqrt{Q_\text{L}}\right)\,\mathrm{sgn}(R)\mathrm{erfc}|R|\,,
\end{multline}
where $R(x,t)\equiv\sqrt{Q_\text{R}}-\sqrt{Q_\text{L}}$ and $\Delta{\hat \mu}={\hat\mu_\text{L}}-{\hat\mu_\text{R}}$.

At generic values of $\xi,$ the difference $R$ is proportional to $\sqrt{Q_\text{R}}-\sqrt{Q_\text{L}}\sim \sqrt t$ implying that the last two lines of Eq. \eqf{jmas} are exponentially suppressed for large $t,$ and only the first line survives. 
This gives a \emph{ballistic} result: both the magnetization density and the current profiles are scaling functions of the  variable $\xi=x/t$, 
and display a jump-like structure  at a critical  ray $\xi = v^*$,   where $Q_\text{R} = Q_\text{L}$, i.e.,  the left and right crossing rates equal
\be
\label{xistar}
\Gamma_\text{R}(v^*)=\Gamma_\text{L}(v^*)\,.
\ee
Using the explicit expressions \eqf{Qdef} this equality can be rewritten in a more illuminating form,
\be
\label{jvn}
j(v^*) = v^* n(v^*)\,,
\ee
where $n(\xi)$ and  $j(\xi)$ are given in Eqs. \eqf{rhop} and \eqf{jp}.
This equation always has a unique solution, which allows us to rewrite the ballistic component as  
\bes
\label{bal}
\begin{align}
m_\text{bal}(\xi) &=n (\xi)\big(\Theta[\xi-v^*]\,{\hat\mu_\text{R}}+\Theta[v^*-\xi]\,{\hat\mu_\text{L}}\big)\,,\label{m_bal}\\
J_\text{bal}(\xi) &= 
j(\xi)
\big(\Theta[\xi-v^*]\,{\hat\mu_\text{R}}+\Theta[v^*-\xi]\,{\hat\mu_\text{L}}\big)\,.\label{j_bal}
\end{align}
\esu

These results have a clear physical meaning.
Recalling the interpretation of $Q_\text{R/L},$ Eq. \eqf{xistar} means that along the ray $v^*$ the fluxes of particles coming from the left and the right are balanced. Equivalently, Eq. \eqf{jvn} 
implies that the fluid velocity defined as $j(x,t)/n(x,t)$ along the ray $x/t=v^*$ is equal to $v^*,$ so in the reference frame traveling at velocity $v^*$ the particle flow is zero.
Thus $v^*$ is the velocity of the boundary between left and right particles. We shall refer to this front of left particles penetrating the gas of right particles or the other way around as the ``second front''. 
The ``first front'' is given by the light cone at $\xi=v_\text{max} =c$ set by the maximal velocity. Inside this light cone the particle current is non-zero.  As the particles carry a finite magnetization, this induces a magnetization current. However, until the second front arrives, i.e. for $v^*<\xi< c,$ its magnitude is set by the average magnetization ${\hat\mu_\text{R}}$ of the right particles, while after the second front has arrived, it is set by the left magnetization ${\hat\mu_\text{L}}.$ For example, in Fig. \ref{sketch} the point $(x,t)$ is inside the light cone but yet to the right of the second front so the spin current is proportional to ${\hat\mu_\text{R}}.$

Does this jump discontinuity in the ballistic result correspond to a physical shock? To answer this, we have to take a closer look at the profiles around the second front at $v^*.$ It is easy to see that around this point it is not justified to drop the terms we neglected in the derivation of Eqs. \eqf{bal}: 
for all $t$ there is a region in $\xi$ around $v^*$ where $R=\sqrt{Q_\text{R}}-\sqrt{Q_\text{L}}$ is small
and the terms we dropped are non-negligible. Expanding around $\xi=v^*$ we find that  
\be
\label{dexp}
R(x,t) = \sqrt{Q_\text{R}} - \sqrt{Q_\text{L}} \approx 
\frac1{2\sqrt{D^*}}
(\xi-v^*)\sqrt{t}\,, 
\ee
with the diffusion constant defined as 
\be
\label{Dstar}
D^*=\frac{\Gamma(v^*)}{n(v^*)^2}\,.
\ee
We thus conclude that the size of the region of $\xi$ where the non-ballistic terms neglected in \eqf{bal} are important shrinks as $\sim t^{-1/2}.$
In terms of the original spacetime variables, however, this corresponds to  a region   $(x-x^*(t))^2 \sim  t D^*$ with  $x^*(t)=v^*t$ denoting 
the instantaneous position of the second front. This shows that the spatial region around the ballistic second front 
actually \emph{grows diffusively} as $\sim\sqrt t$ and  there is no real shock wave. 

Let us now focus to the region close to the ballistic second front by taking the limits 
 $t\to\infty$ and $R=\sqrt{Q_\text{R}}-\sqrt{Q_\text{L}}$ fixed. Then  the last two lines of Eq. \eqf{jmas} can be simplified further,
\begin{multline}
\label{jmerf}
\begin{pmatrix}
J(x,t)\\
m(x,t)
\end{pmatrix}
 \approx \begin{pmatrix}j (\xi)\\ n(\xi)\end{pmatrix}
 \big(\Theta[\xi-v^*]{\hat\mu_\text{R}}+\Theta[v^*-\xi]{\hat\mu_\text{L}}\big)\\
+\begin{pmatrix}j(v^*)\\n(v^*)\end{pmatrix}
\frac{{\hat\mu_\text{L}}-{\hat\mu_\text{R}}}2 \mathrm{sign}(\xi-v^*)\,
\mathrm{erfc}\left(\frac{|\xi-v^*|\sqrt{t}}{\sqrt{4D^*}}\right)\,.
\end{multline}

This equation provides a surprisingly accurate approximation for the exact magnetization profiles, Eq.~\eqref{jmres} at large times. 
In Figs. \ref{fig:mprofiles} and \ref{fig:jprofiles}  we plot the magnetization and the spin current using thermal momentum distributions and relativistic dispersion relation (see caption of Fig. \ref{fig:mprofiles} for details). Fig. \ref{fig:jprofiles}.a demonstrates that the asymptotic expression \eqf{jmerf} plotted in dashed line approaches the full result \eqf{jmres} shown as a solid line. A similar behavior is observed for the magnetization profiles (not plotted in Fig. \ref{fig:mprofiles}.a). The ballistic solutions \eqf{bal} with the jump discontinuity are plotted in Figs. \ref{fig:mprofiles}.b and \ref{fig:jprofiles}.b in dashed line.
Results of the Monte Carlo simulations discussed in Section~\ref{sec:level1} are shown as symbols.

Both the $\sqrt{t}$ dependence and the appearance of the error function hints at the diffusive nature of the correction. Indeed, around the second front $\xi\approx v^*,$ the magnetization is
\be
\label{mfront}
m(x,t) \approx 
n^*\; \frac{{\hat\mu_\text{L}}+{\hat\mu_\text{R}}}2 - n^*\; \frac{{\hat\mu_\text{L}}-{\hat\mu_\text{R}}}2 \mathrm{erf}\left(\frac{x-x^*(t)}{\sqrt{4D^*t}}\right)\,.
\ee
In the reference frame of the front, this is just the solution of the diffusion equation with diffusion constant $D^*$ with step-like initial condition. Our formula thus describes the \emph{diffusive broadening of the ballistically moving front}.

The NESS is obtained by setting $\xi=0,$ $t\to\infty,$ yielding (for $v^*\neq0$)
\be
\label{NESS}
\begin{pmatrix}
J_\text{NESS}\\
m_\text{NESS}
\end{pmatrix}
 = \begin{pmatrix}j(0)\\
n(0)\end{pmatrix}
 \big(\Theta[-v^*]{\hat\mu_\text{R}}+\Theta[v^*]{\hat\mu_\text{L}}\big)\,.
\ee
Thus for right moving (left moving) fronts  the magnetization and its current in the NESS are determined   by the polarization on the left (right)
and the average densities and particle currents of particles passing through the origin (see Eqs.~\eqref{rhop} and \eqref{jp}). 

Interestingly, we can also compute the magnetization density and current analytically right at the front using that the last term in Eq. \eqf{jmres} vanishes and $\int_1^\infty \ud z \frac{e^{-2Q^*z}}{z} I_1\left(2Q^*\,z\right)=e^{-2Q^*} [I_0(2Q^*)+I_1(2Q^*)]$ 
with the result
\bes
\label{atfront}
\begin{align}
\begin{split}
m|_{x/t=v^*} &= \frac{{\hat\mu_\text{L}}+{\hat\mu_\text{R}}}2 n(v^*) \\
-& \frac{{\hat\mu_\text{L}}-{\hat\mu_\text{R}}}2(\p_x{Q}^*_\text{R}+{\p_xQ}^*_\text{L})e^{-2Q^*}I_0(2Q^*)\,,
\end{split}\\
\begin{split}
J|_{x/t=v^*} &= \frac{{\hat\mu_\text{L}}+{\hat\mu_\text{R}}}2 j(v^*) \\
+& \frac{{\hat\mu_\text{L}}-{\hat\mu_\text{R}}}2(\p_t Q^*_\text{L}+\p_t Q^*_\text{R})e^{-2Q^*}I_0(2Q^*)\,,
\end{split}
\end{align}
\esu
where we used the notation $Q^*(t)=\Gamma(v^*)\,t.$ The time dependence comes from the factors
$e^{-2P^*t}I_0(2P^*t)$ that 
give a $\sim 1/\sqrt t$ approach towards the large time asymptotic values  $m(v^*)(\hat\mu_\text{L}+\hat\mu_\text{R})/2$ and $j(v^*)(\hat\mu_\text{L}+\hat\mu_\text{R})/2.$

\subsection{Balanced case: diffusive spin transport}

The asymptotic result in Eq. \eqf{jmerf} is not correct for the special, balanced case, when the streams of particles coming from the left and right balance each other such that the front does not move, $v^*=0$. In this case we have   $\Gamma_\text{L}(v^*)=\Gamma_\text{R}(v^*)$, which, through the general 
relation  $\p_t Q(v^*) = \Gamma(v^*)-v^* \p_x \Gamma(v^*)$  immediately yields that  
  $\p_t Q_\text{R}(v^*)=\p_t Q_\text{L}(v^*)$, i.e. that the particle current at the front vanishes $j(v^*)=0$. 
 Apart from fine tuned cases, this can happen most naturally in a balanced situation when $f_\text{L}(p) = f_\text{R}(p)  = f(p) = - f(p),$  i.e. when the even momentum distributions 
and the average densities on the two sides are equal, only the spin distributions are different. Then the particle current and the associated  ballistic component of the magnetization current is identically zero,  $\p_t Q_\text{R} - \p_t Q_\text{L}=0,$ and $\p_x Q_\text{R}-\p_x Q_\text{L} = n$ so the 
orbital degrees of freedom are homogeneous throughout the system. 
For thermal initial states this is the case when $T_\text{L}=T_\text{R}$ and $h_\text{L}=-h_\text{R}$ implying opposite magnetizations ${\hat\mu_\text{R}}=-{\hat\mu_\text{L}}.$

In this balanced case the magnetization dynamics is entirely described by the diffusive component. At the origin we find, in particular
\begin{align}
Q_\text{R}(0)=Q_\text{L}(0)&=t\int\frac{\ud p}{2\pi} \Theta(v_p)f(p)v_p=t/(2\tau)\,,\\
\p_t Q_\text{R}(0)=\p_t Q_\text{L}(0) &= \int\frac{\ud p}{2\pi} \Theta(v_p)f(p) v_p=1/(2\tau)\,,\\
\p_xQ_\text{R/L} (0)&= \pm\int\frac{\ud p}{2\pi} \Theta(\mp v_p)f(p)=\pm n/2\,,
\end{align}
where we introduced the collision time $\tau$ defined as the ratio of the average separation and the average velocity modulus,
\be
\label{taudef}
\tau^{-1} =  n {\vev{|v|}} = \int \frac{\ud p}{2\pi} f(p) |v_p|  \,.
\ee
Using Eqs. \eqf{atfront} we find that at the origin 
\begin{align}
m(0,t) &= n \frac{{\hat\mu_\text{L}}+{\hat\mu_\text{R}}}2\,,\\
J(0,t) &= 
\frac{\Delta{\hat \mu}}2\frac{e^{-t/\tau} }{\tau} I_0 (t/\tau)\,.
\end{align}
After a sudden jump, the magnetization current  decreases linearly for short times,
$J(0,t) \approx\Delta{\hat \mu}/(2\tau)\,(1-t/\tau),$
while for large times it decays to zero,
\be
J(0,t)\approx \frac{\Delta{\hat \mu}}{\sqrt{2\pi t \;\tau}}\,.
\label{eq:J(0t)_diffusive}
\ee
Both the short and the long time behavior are governed by the collision time $\tau.$

Expanding around the origin then gives
$\sqrt{Q_\text{R}}-\sqrt{Q_\text{L}}\approx x/(2\sqrt{Dt}),$
where the diffusion constant  is  proportional to the collision rate, Eq.~\eqref{taudef}. 
\be
\label{Dbal}
D=\frac1{2 n^2\tau} %= \frac{\int \frac{\ud p}{2\pi} f(p) |v_p| }{2\left(\int \frac{\ud p}{2\pi} f(p) \right)^2}\,.
\ee

The magnetization density is given by Eq. \eqf{jmerf}, while the current can be obtained from Eq. \eqf{jmas}. Now due to $\p_t Q_\text{R}=\p_tQ_\text{L}$ an extra factor of $Q_\text{L}-Q_\text{R}$ appears in the numerator of the last term rendering the last two lines of the same order. Luckily, a cancellation takes place and we find  
\bes
\label{balanced}
\begin{align}
m(x,t) &\approx
n\frac{{\hat\mu_\text{L}}+{\hat\mu_\text{R}}}2 - n\frac{{\hat\mu_\text{L}}-{\hat\mu_\text{R}}}2 \mathrm{erf}\left(\frac{x}{\sqrt{4Dt}}\right)\,,
\\
J(x,t) &\approx n\frac{\Delta{\hat \mu}}2\sqrt{\frac{D}{\pi t}} e^{-x^2/(4Dt)}\,.
\end{align}
\esu
The magnetization profile for large times is the solution of the diffusion equation
\be
{\p_t m(x,t)} = D\; {\p_x^2 m(x,t)} \,,
\ee
with the step-like initial condition $m(x,0)=\Theta(-x)\,n\,{\hat\mu_\text{L}}+\Theta(x)\,n\,{\hat\mu_\text{R}}.$

We note that in the specific case of the $O(3)$ non-linear sigma model with thermal non-relativistic momentum distribution, $f(p)=[1+2\cosh(\beta h)]e^{-\beta\eps(p)},$ the diffusion constant \eqf{Dbal} coincides with that extracted from the thermal dynamical spin-spin correlation function using the semiclassical method in Ref. \cite{Sachdev1997}.

\section{Beyond the universal limit}

The analytic treatment discussed in the previous sections relied on the assumption of
 purely reflective scattering, valid in the limit of vanishing  quasiparticle momenta. 
For faster quasiparticles, however, transmissive processes are also allowed, moreover, 
the set of outgoing spins need not coincide with the incoming set.
In this section, we shall investigate the effect of these non-transmissive processes 
through  Monte Carlo simulations \cite{Moca2016}. 
For simplicity, we  focus on the O(3) $\sigma$-model, where 
the magnetization is locally conserved, so the $S$-matrix has non-zero matrix elements between states of the same total spin component in the $z$ direction. In a collision of a $+$ and a $-$ particle 
there are three possible out states, $(+,-)\longrightarrow (+,-),$ $(-,+),$ $(0,0),$ corresponding to reflection, transmission, and transmutation into $\mu=0$ particles.

In the simulations we  average over  semiclassical configurations numerically. %
While generating the initial quasiparticle configurations 
and finding the coordinates of the collisions is simple, following the trajectories of the spins is more complicated. In the universal limit, spins get reflected at each  collision, and follow  zig-zag paths. 
Averaging the snapshots of the spin positions at given times over the semiclassical configurations yields the 
 density profiles for each particle species, and   determines the magnetization profile. 
 This is how the numerical data shown in Figs. \ref{fig:mprofiles}, \ref{fig:jprofiles} was obtained using two particle species, i.e. when $\mu$ can take two values $\mu=\pm1.$

In the non-universal case, different spin configurations are superposed  with 
the respective probability amplitudes of the 2-particle S-matrix. 
The issuing coherent time evolution of the quasiparticle spins   can be described by a hybrid semiclassical
matrix product state approach~\cite{Moca2016}, whereby  
 the spatial sequence of the particle spins is treated  as an effective spin chain. 
  Here, however,  we focus on magnetization densities, and we do not expect spin coherence 
  to play a major role \footnote{Indeed, the two versions of the hybrid method was found to give identical results even for equal time correlation functions after a quantum quench.}.
Therefore we use a simpler, \emph{classical}  version of the hybrid method where we disregard the quantum coherence of spins and  assign  specific outcomes of  each collision with  probabilities derived from the respective 
  $S$-matrix amplitudes, $|S_{\mu_1\mu_2}^{\mu_1'\mu_2'}|^2$.

\begin{figure}[t!]
\includegraphics[width=.5\textwidth]{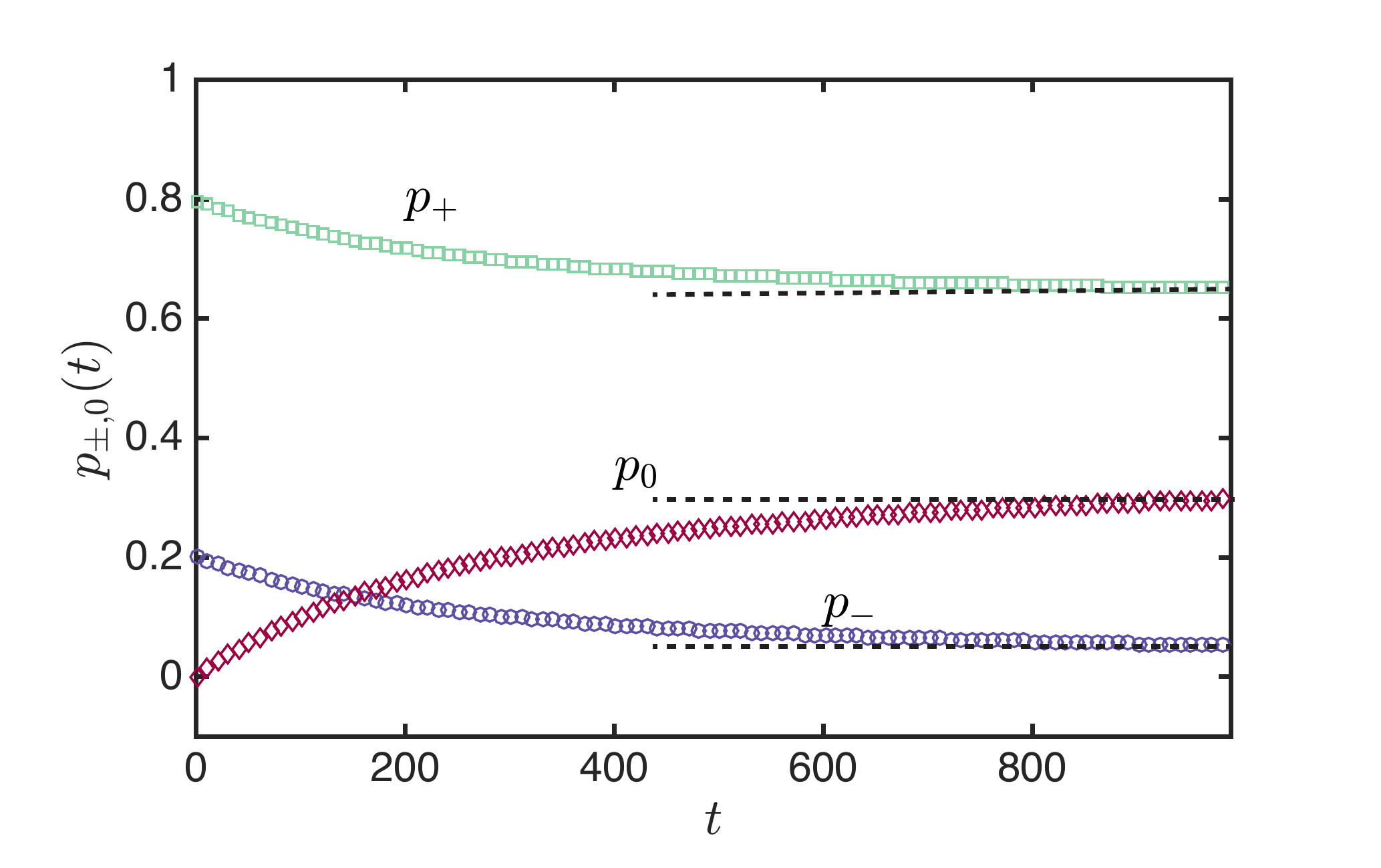}
\caption{\label{fig:equil}(a) Time evolution of the relative densities the three particle species in the homogeneous $O(3)$ non-linear sigma model. Initially, $p_+^{(0)}=0.8,$ $p_-^{(0)} = 0.2,$ $p_0^{(0)}=0.$ The velocity distribution of the particles is thermal with inverse temperature $\Delta \,\beta = 2$ where $\Delta$ is the particle gap. Time $t$ is measured in units of $\hbar/\Delta$ with $\hbar$ set to $1.$ Dashed lines indicate the values $p_+=0.66,$ $p_-=0.06,$ $p_0=0.28$ obtained 
from the detailed balance condition, Eq.~\eqref{detbal}.}
\end{figure}

\label{sec:level1}

\subsection{Relaxation of particle densities in a homogeneous system}

\label{sec:homog}

Let us study first the relaxation of the number (density) of particles with a given spin in a spatially {\em homogeneous} setup. Though the total quasiparticle number $N$ and the spin $N_+-N_-$ are both conserved, 
$(0,0)\leftrightarrow (+,-)$ scattering leads to a relaxation of the particle 
numbers $N_\mu$ and the corresponding  densities $n_\mu$ and probabilities 
$p_\mu=N_\mu/N.$

 \begin{figure}[t!]
	\includegraphics[width=.5\textwidth]{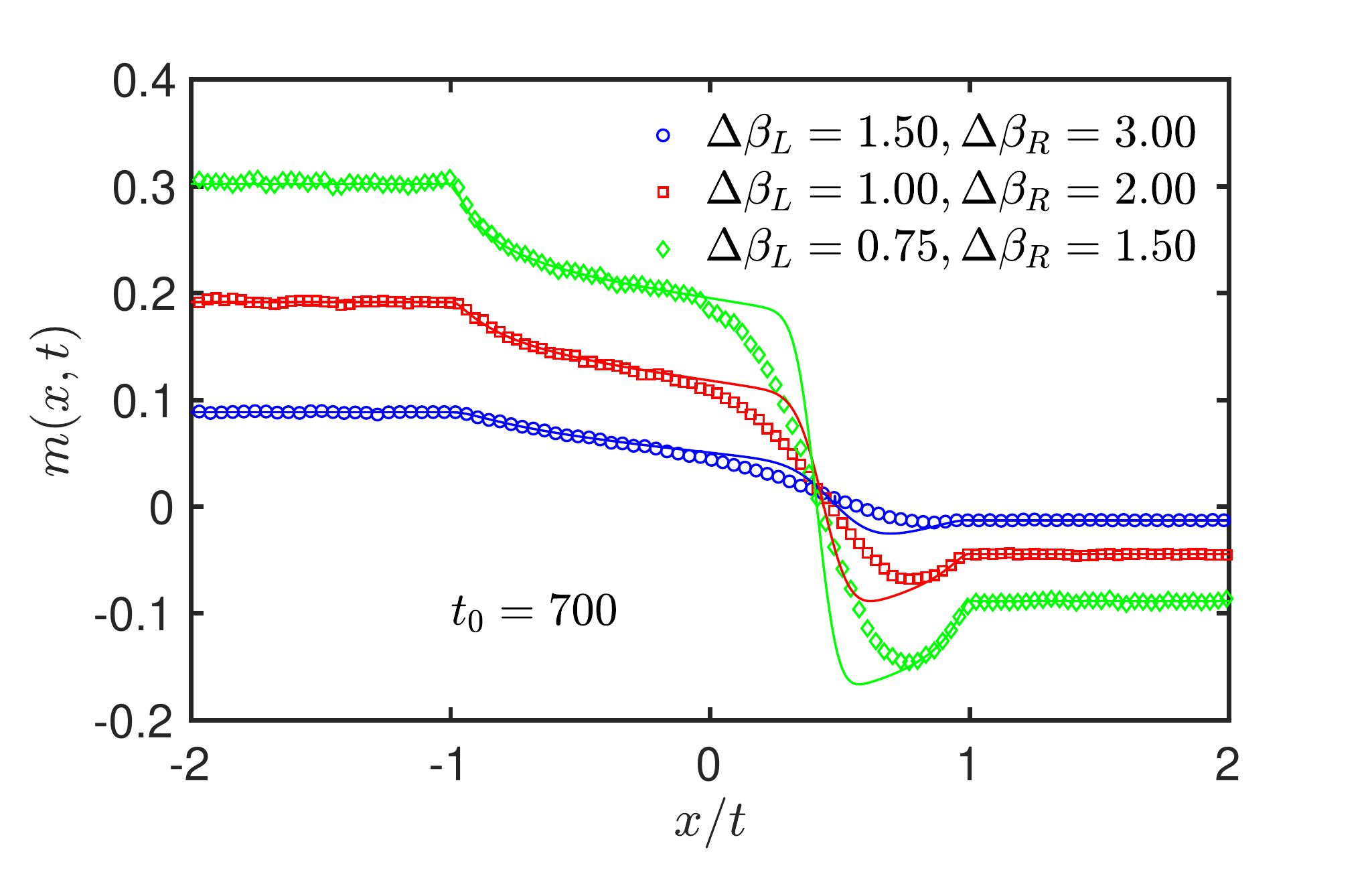}
	\caption{\label{fig:profiles_lev1}  Magnetization profiles in the non-universal limit at time $\Delta\,t=700$ 
		plotted as functions of $x/t$ for left/right momentum distributions $ f_\text{L/R}(p)\sim e^{-\beta_\text{L/R}\eps(p)}$ with different inverse temperatures as indicated in the legend and average magnetizations per particle ${\hat\mu_\text{L}}=1,$ ${\hat\mu_\text{R}}=-1.$ 
%The ratio $\beta_\text{R}/\beta_\text{L}$ is fixed to 2.
The solid lines are the analytical results in the universal limit given by Eq.~\eqf{jmres}. The coordinate and time are measured in units of $\Delta$ as in Fig. \ref{fig:mprofiles} and we set $\hbar=c=1.$}
\end{figure}

\begin{figure}[t!]
	\includegraphics[width=.45\textwidth]{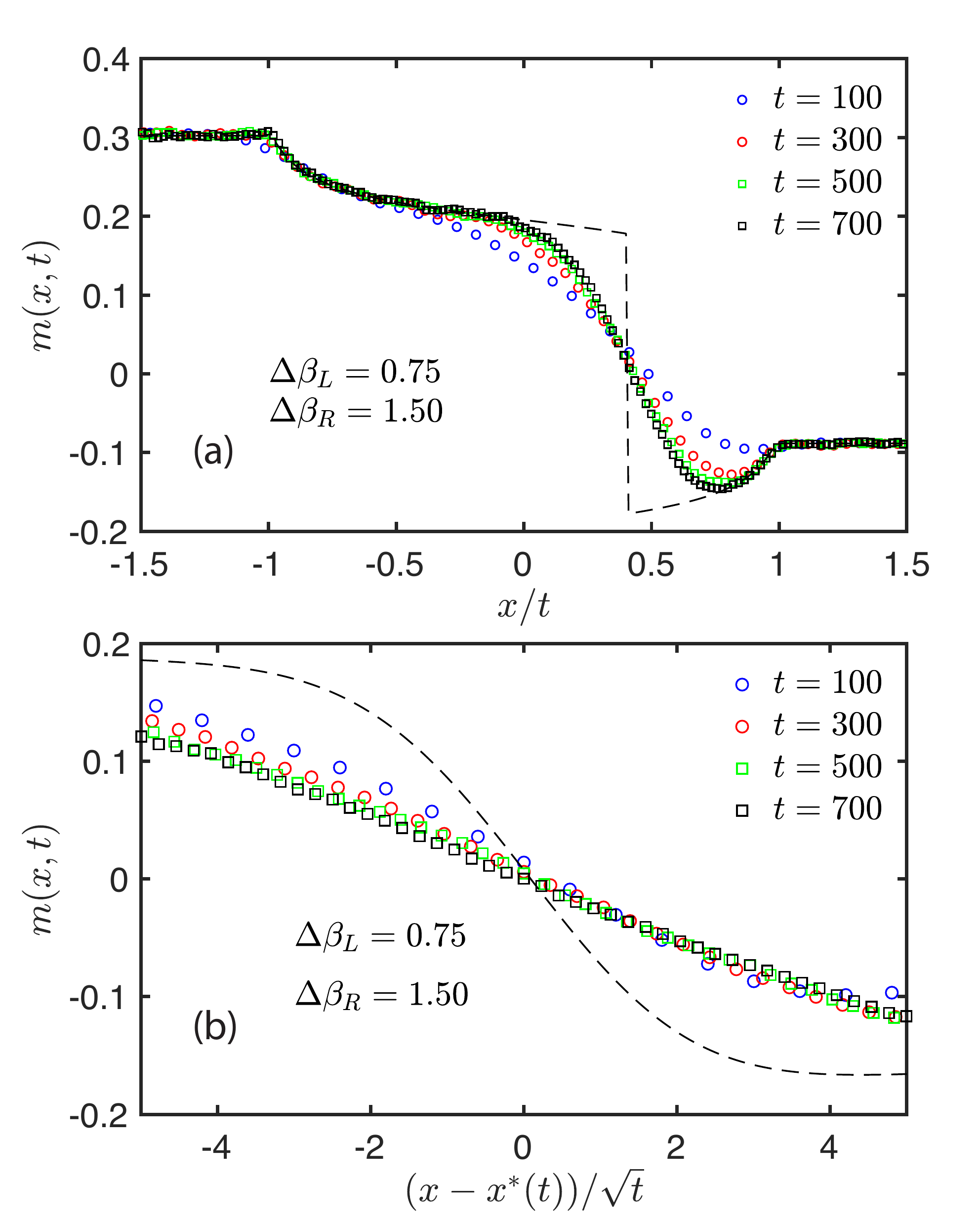}
	\caption{\label{fig:mag_lev1} (a) Magnetization profiles in the non-universal case at different times as indicated in the legend plotted as functions of $x/t$ for temperatures $\Delta\,\beta_\text{L}=0.75,$ $\Delta\,\beta_\text{R}=1.5$ and average magnetizations per particle ${\hat\mu_\text{L}}=1,$ ${\hat\mu_\text{R}}=-1.$ 
		 The ballistic result, Eq.~\eqf{m_bal}, in the universal limit 
		is represented by a dashed line.  (b)  Rescaled magnetization profiles around the second front  as a function of $[x-x^*(t)]/\sqrt{t}$, demonstrating the diffusive broadening of the front. The dashed line represents the 
		analytical result for the diffusive behavior in the universal (reflective) limit. 
	}
\end{figure}

\begin{figure}[t!]
\includegraphics[width=.45\textwidth]{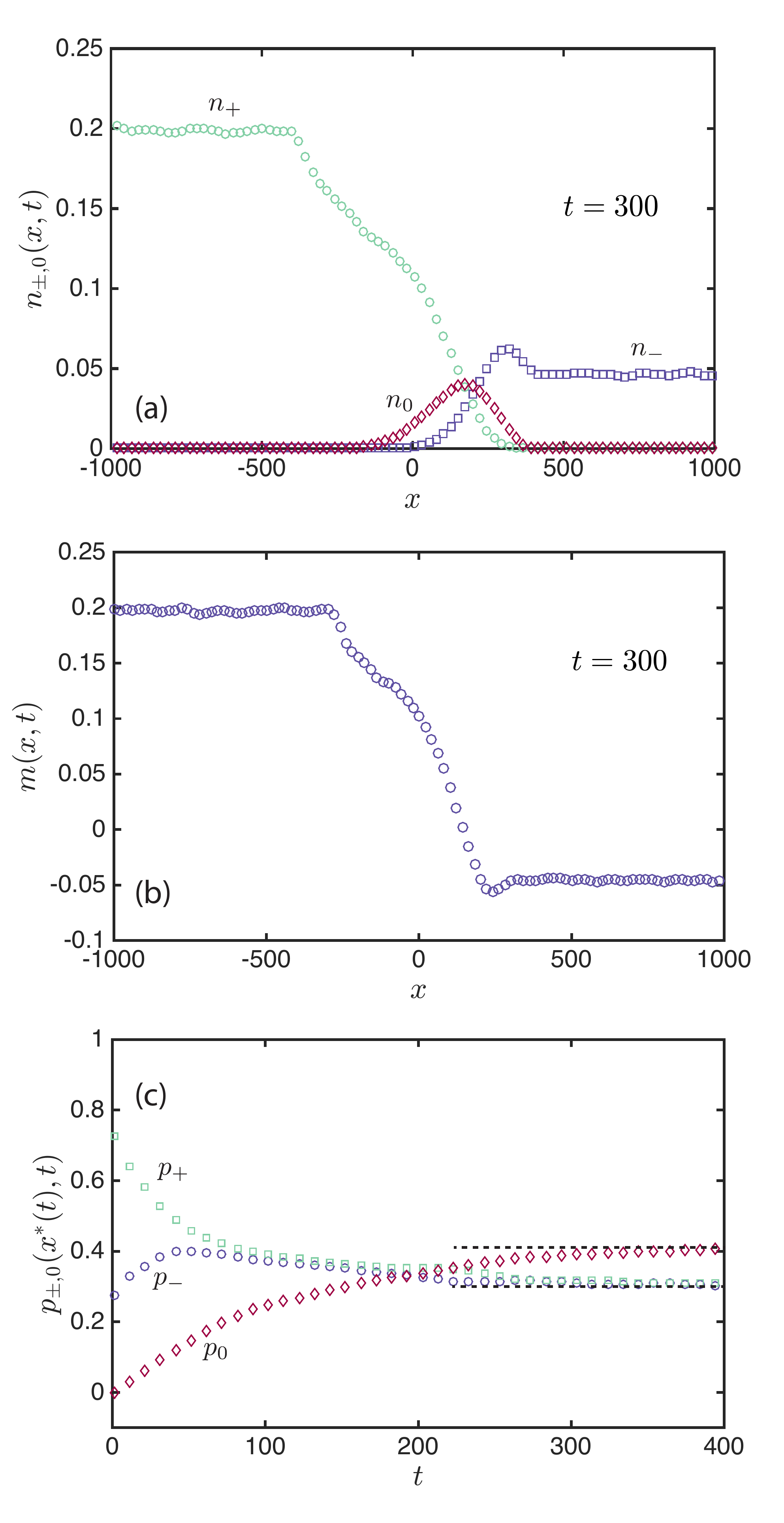}
\caption{\label{fig:frontequil} (a) Density profiles of the particle species with $S^z=+1,$ $-1,$ and $0$ in the $O(3)$ non-linear sigma model at time $t/(\hbar/\Delta)=300$ where $\Delta$ is the mass gap. The coordinate $x$ is measured in units of the Compton length $\hbar c/\Delta.$ The initial left/right momentum distributions are $ f_\text{L/R}(p)=e^{-\beta_\text{L/R}\eps(p)}$ with $\eps(p)=\sqrt{\Delta^2+p^2c^2},$ inverse temperatures $\Delta\,\beta_\text{L}=1,$ $\Delta\,\beta_\text{R}=2.$ The initial state is fully polarized, i.e.  the average magnetizations per particle are ${\hat\mu_\text{L}}=+1,$ ${\hat\mu_\text{R}}=-1.$ (b) Magnetization density profile in the same case.
(c) Time evolution of the relative densities of the particle species near the second front. Time is measured in units of $\hbar/\Delta.$ Dashed lines indicate the values obtained 
from the detailed balance condition, Eq.~\eqref{detbal}.}
\end{figure}

The time evolution of the occurrences $p_\mu(t)$  is shown in Fig.~\ref{fig:frontequil}
for an initial state of thermalized quasiparticles with no $\mu=0$ particles and $80\%$ ($20\%$)  of type $+$  ($-$)
particles. 
By  parity and time reversal, however, 
the equilibrium densities \emph{must}  satisfy detailed balance,  
\be
\label{detbal}
p_+\,p_- = \frac12\; p_0^2\,,
\ee

where the factor $1/2$ takes into account that  colliding  $0$ particles are identical. Noticing that 
only  collisions of type $(0,0)\leftrightarrow (+,-)$ generate particle number  relaxation, 
we  can parameterize the occurrences of the particle species as 
$p_\pm = p_\pm^{(0)}-x$ and  $p_0=p_0^{(0)}+2x$. Substitution to  Eq.~\eqref{detbal} 
then determines the stationary values of the probabilities $p_\mu$. 
Indeed, the $p_a(t)$ curves in Fig. \ref{fig:equil}  approach these values, 
clearly demonstrating that the hybrid semiclassical method is able to capture  
equilibration of particle species.

\begin{figure}[t!]
\includegraphics[width=.45\textwidth]{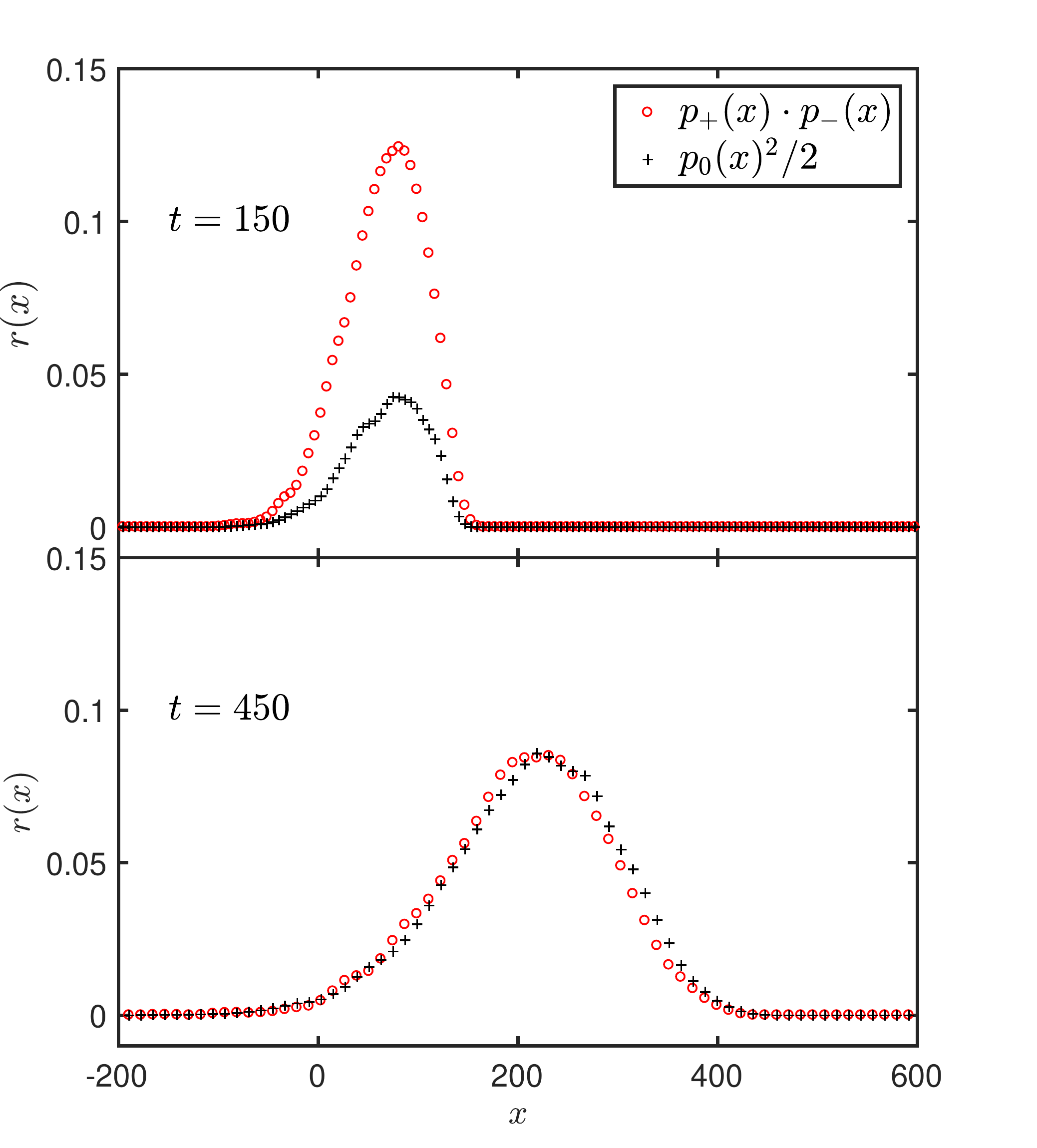}
\caption{\label{fig:frontequil2} The left and right hand side of the detailed balalnce condition \eqf{detbal} involving the relative populations of the three particle species as functions of the position for the setup initial to that in Fig. \ref{fig:frontequil}. The dynamics leads to local equilibration: at shorter times (top panel) detailed balance is clearly violated. However, at later times (bottom) local equilibration takes place, and the local densities satisfy detailed balance at each point. Units and conventions as in Fig. \ref{fig:frontequil}.}
\end{figure}

\subsection{Transport beyond the universal limit}

Let us now turn to the setup studied in Sections \ref{sec:deriv} and \ref{sec:asym} to see the effect of the non-universal S-matrix on the non-equilibrium dynamics and transport. Here we focus on the fully polarized initial state with 
 only $+$ particles on the left, and only $-$ particles on the right.
 
 First we check whether the scaling properties of the profiles change upon allowing for non-trivial scattering. In Fig. \ref{fig:profiles_lev1} we plot the snapshot of the magnetization profiles at a given time for three different left/right temperatures together with the ones corresponding to the universal limit. It is clear that the non-trivial S-matrix has a significant effect but mainly around the second front. As expected, the front broadens more rapidly as there is a finite probability of pure transmission at collisions. In Fig. \ref{fig:mag_lev1}.a the magnetization profile is shown for different times as a function of $x/t$ to demonstrate that, similarly to the universal case, the curves collapse on the universal ballistic profile far from the second front. The behavior around the front is shown in Fig. \ref{fig:mag_lev1}.b demonstrating that the width of the front grows as $\sqrt t$ so it still broadens diffusively, as expected, albeit with a different (larger) diffusion constant. In summary, the picture of a ballistically propagating and diffusively spreading ``second front'' remains valid also in the non-universal case.

Next we analyze the effect of the flavor changing scattering processes on the densities of the different particle species to see if a relaxation similar to that found in the homogeneous case in Sec. \ref{sec:homog} takes place also in the inhomogeneous setup.
 Fig. \ref{fig:frontequil}.a shows  the density profiles of the three particle species separately. The second front moving with  velocity $x/t=v^*$ is the location of the boundary between the $+$ and $-$ particles coming from the left and from the right, respectively. The magnetization profile given by $m(x,t)=n_+(x,t)-n_-(x,t)$ is shown in Fig.~\ref{fig:frontequil}.b. As we saw, the qualitative behavior of the magnetization and the spin current is the same as in the universal case. However, due to the collisions of $+$ and $-$ particles, a bump of $\mu=0$ particles develops around the front as shown in Fig. \ref{fig:frontequil}. a. 

As time evolves, {\em local equilibration} takes place around the front. In Fig. \ref{fig:frontequil}.c we show the populations of the particle species around the front at $x = v^* t.$ 
The analytical result,  Eq.~\eqf{mfront} predicts asymptotically a magnetization  $m(x/t=v^*) = n^*(\hat\mu_\text{L}+\hat\mu_\text{R})/2$ so $p_+-p_- = (\hat\mu_\text{L}+\hat\mu_\text{R})/2$ at the front. 
Assuming that this result carries over to  non-reflective scattering,
we can use again the detailed balance condition \eqf{detbal} 
to determine the proportion of charged particles at the front. 
In Fig.~\ref{fig:frontequil} we have  $(\hat\mu_\text{L}+\hat\mu_\text{R})/2=0$, yielding 
$p_{\pm}  = \frac{2-\sqrt2}2\approx0.293$ and $ p_0=\sqrt2-1\approx0.414$.
These values are in very good agreement with the numerics, 
 shown in Fig. \ref{fig:frontequil}.c, thus demonstrating local equilibration of particles with different spins at the second front.

This local equilibration takes place not only at the second front but in a region around it, and eventually in the whole system. In Fig. \ref{fig:frontequil2} we show the quantities at the two sides of the detailed balance condition \eqf{detbal} as functions of the position $x$ for two different times. For the fully polarized initial state, 
the densities on the left and right hand side satisfy the detailed balance 
condition \footnote{For initial states not obeying detailed balance there is a nontrivial evolution towards local equilibration even far from the second front.}. 
Near the second front where the left and right particles meet, detailed balance is not satisfied initially. 
However,  local equilibration takes place with time, and detailed balance is restored.

\section{Conclusions}
\label{sec:concl}

In this work we used the semiclassical approach to study transport and front propagation 
in systems with massive 
``spinful"  quasiparticles with an internal quantum number.
We found that unlike the energy and density that propagate ballistically, the dynamics of internal degrees of freedom is generically not ballistic (unless the scattering of particles is  completely transmissive), but rather ballistic front propagation and diffusive dynamics coexist.

In the universal low energy limit of fully reflective scattering, in particular,  we derived analytical expressions for the evolution of the magnetization density and spin current profiles for the case of two semi-infinite systems joined at time $t=0.$
We find that spin transport can be diffusive or ballistic depending on the initial state. A purely diffusive behavior arises in the `balanced' case, when the densities and momentum distributions are the same on the two sides, and only spin polarizations differ. This is  reminiscent to the situation studied in Ref. \cite{Ljubotina2017} where diffusive behavior was observed numerically in the XXZ spin chain for equal temperatures and opposite magnetic fields on the two sides.

In the  -- more generic -- imbalanced case, magnetization density and spin current profiles have a jump discontinuity as a function of $\xi=x/t$ in the 
limit $x,t\to\infty$ with $\xi$ finite. This jump corresponds to the interface between left and right particles. This is somewhat similar to, but not  the same as the jumps predicted by the GHD framework in the gapped phase of the XXZ spin chain located at the maximal velocities of the different quasiparticle species \cite{Piroli2017}. Remarkably and in contrast to the GHD description \cite{Fagotti2017}, our analytic expressions capture 
also the sub-ballistic, diffusive broadening of this front.

To go beyond the universal low momentum regime, we also studied the effect of non-reflective scattering on  transport and front propagation phenomena in the $O(3)$ non-linear sigma model using a  hybrid semiclassical Monte Carlo approach~\cite{Moca2016}. This approach accounts also for processes where  individual spins are not conserved in the collisions, so that  populations of the three possible spin states can change in time. Interestingly, we find that  spin populations reach local equilibrium rather quickly, and the propagating  front obeys 
local \emph{detailed balance}. However, the diffusively broadening, ballistically propagating second front in the magnetization density is still present and appears to be a generic feature of inhomogeneous systems with internal degrees of freedom, at least in the semiclassical regime.

Finite temperature spin transport in the $O(3)$ non-linear sigma model has been studied in the past in the context of dynamical spin current correlation functions. While calculations based on the thermodynamical Bethe Ansatz \cite{Fujimoto1999} and form factor expansions \cite{Konik2001} found a finite spin Drude weight at zero magnetic field, the semiclassical approach predicts vanishing Drude weight\footnote{In this context, the Drude weight refers to a $\delta(\omega)$ peak in the spin conduction. In contrast, semiclassics yields a Drude peak of finite width, $\omega \sim 1/\tau,$ just as in the Drude theory of metals.}. Resolving this discrepancy is beyond the scope of our paper, but let us point out that the connection between the nature of front broadening and thermal dynamic correlation functions is not direct. For example, in the case of free fermions, a clearly ballistic system, the front shows a universal subdiffusive $\sim t^{1/3}$ broadening \cite{Hunyadi2004,Eisler2013,Zauner2015}.

From the viewpoint of Bethe Ansatz, our model belongs to the class of systems having non-diagonal scattering. The first application of GHD in such a system was done in the recent contribution \cite{Ilievski2017}, but further implications of the non-diagonal nature of the scattering remain to be studied. 
We believe that our semiclassical calculations  provide valuable benchmarks for the further development of hydrodynamic descriptions.

Our system can also  be related to the so-called classical soliton gas that was proposed to provide a physical picture for the GHD equations \cite{Doyon2017e}. 
It would be interesting to include velocity dependent time delays of colliding particles, classical counterparts of quantum scattering phase shifts, that are 
essential for reproducing the structure of the GHD equations \cite{Doyon2017e}.

The semiclassical approach applied here has many perspectives.
It can and has been used to describe the time evolution of correlation functions \cite{Rieger2011,Evangelisti2013,Kormos2015,Moca2016}, carrying valuable information in the current setup as well \cite{Doyon2017f}.  These calculations could possibly be extended to inhomogeneous non-equilibrium states, 
investigated here.  It may also be possible to extend  the semiclassical description to higher dimensions. Here collisions cannot be treated in terms of point-like particles but a finite cross section must be introduced. Moreover, different geometries may lead to different types of behavior as the particle density at the front can change in time. 
We leave these interesting questions and directions for future study.

{\em Acknowledgements.} We gratefully thank Spyros Sotiriadis and Toma\v z Prosen  for fruitful discussions.  This work was supported by 
 the National Research Development and Innovation Office of Hungary within the Quantum Technology National Excellence Program  (Project No. 2017-1.2.1-NKP-2017-00001) and under OTKA grant No. SNN118028. M.K. was partially supported by NKFIH
K-2016 grant no. 119204 and a Pr\'emium Postdoctoral Fellowship of the HAS.  C.P.M. was supported by the Romanian National Authority for Scientific Research and Innovation, UEFISCDI, project number PN-III-P4-ID-PCE-2016-0032.

\newpage

\bibliographystyle{apsrev4-1}

\bibliography{profile_paper2col_v6}

\setcounter{section}{0}
\renewcommand{\thesection}{A\arabic{section}}
\renewcommand{\theequation}{A\arabic{equation}}
\setcounter{equation}{0}%

\onecolumngrid

\section*{Appendix}

\section{Details of the derivation of Eq. \eqf{RES1}}
\label{app:deriv}

Let us first compute $\vev{\Theta(s)s}$ where $s$ is given in Eq. \eqf{s}, using formula \eqf{ave}.
In order to decouple the coordinates of the different particles, we employ the integral representation for the Heaviside theta function,
\be
\Theta(x) = \int \frac{\ud u}{2\pi} \frac{e^{iux}}{iu+\eps}\,,
\ee
and obtain
\be
\begin{split}
\vev{\Theta(s)s}= &  
\frac1{N_\text{R}^{N_\text{R}}}\prod_{i=1}^{N_\text{R}}\int_0^L\ud y_i \int \frac{\ud p_i}{2\pi} f_\text{R}(p_i)
\frac1{N_\text{L}^{N_\text{L}}}\prod_{j=1}^{N_\text{L}}\int_{-L}^0\ud \bar y_j \int \frac{\ud \bar p_j}{2\pi} f_\text{L}(\bar p_j)\\
& \int \frac{\ud u}{2\pi} \frac1{iu+\eps}e^{iu\sum_{j=1}^{N_\text{L}}\Theta(\bar y_j+\bar v_jt-x) - iu\sum_{j=1}^{N_\text{R}}\Theta(x-y_j-v_jt)}
\left[\sum_{j=1}^{N_\text{L}}\Theta(\bar y_j+\bar v_jt-x) - \sum_{j=1}^{N_\text{R}}\Theta(x-y_j-v_jt)\right]
\,,
\end{split}
\ee
where $v_j=v_{p_j}$ and $\bar v_j=v_{\bar p_j}$
Now the multiple integral over positions and velocities can be factorized, leading to double integrals like 
\begin{subequations}
\begin{align}
\frac1{N_\text{R}}\int_0^L\ud y \int  \frac{\ud p}{2\pi} f_\text{R}(p) e^{-iu\Theta(x-y-v_pt)} \Theta(x-y-v_pt)= 
\frac1{N_\text{R}}\int \frac{\ud p}{2\pi}f_\text{R}(p)\Theta(x/t-v_p)   (x-v_pt) e^{-iu}
&=\frac{Q_\text{R}}{N_\text{R}}e^{-iu}\,,\\
\frac1{N_\text{R}}\int_0^L\ud y \int  \frac{\ud p}{2\pi}f_\text{R}(p) e^{-iu\Theta(x-y-vt)} 
=1 + \frac1{N_\text{R}}\int  \frac{\ud p}{2\pi}f_\text{R}(p) \Theta(x/t-v_p) (x-v_pt)(e^{-iu}-1)&=1+\frac{Q_\text{R}}{N_\text{R}}(e^{-iu}-1)\,,
\end{align}
\end{subequations}
where $Q_\text{L/R}$ are defined in Eq. \eqf{Qdef}. Evaluating the other integrals in a similar manner we obtain
\begin{multline}
\vev{\Theta(s)s}=
\int \frac{\ud u}{2\pi} \frac1{iu+\eps}\left\{
N_\text{L} \frac{Q_\text{L}}{N_\text{L}}e^{iu}\left[1+\frac{Q_\text{L}}{N_\text{L}}(e^{iu}-1)\right]^{N_\text{L}-1}\left[1+\frac{Q_\text{R}}{N_\text{R}}(e^{-iu}-1)\right]^{N_\text{R}}\right.\\
-\left.N_\text{R} \frac{Q_\text{R}}{N_\text{R}}e^{-iu}\left[1+\frac{Q_\text{R}}{N_\text{R}}(e^{-iu}-1)\right]^{N_\text{R}-1}\left[1+\frac{Q_\text{L}}{N_\text{L}}(e^{iu}-1)\right]^{N_\text{L}}
\right\}\,.
\end{multline}
In the thermodynamic limit $N_\text{R},N_\text{L},L\to\infty$ with the densities $n_\text{R/L}=N_\text{R/L}/L$ fixed, so 
\be
\begin{split}
\vev{\Theta(s)s}&=
\int \frac{\ud p}{2\pi} \frac1{iu+\eps}
\left(Q_\text{L}e^{iu}e^{(e^{iu}-1)Q_\text{L}}e^{(e^{-iu}-1)Q_\text{R}}-Q_\text{R}e^{-iu}e^{(e^{-iu}-1)Q_\text{R}}e^{(e^{iu}-1)Q_\text{L}}\right)
\\
&=2\sqrt{Q_\text{R}Q_\text{L}}e^{-Q_\text{R}-Q_\text{L}}\int \frac{\ud u}{2\pi} \frac1{u-i\eps}\sin(u-i\gamma) e^{2\sqrt{Q_\text{R}Q_\text{L}}\cos (u-i\gamma)}\,,
\end{split}
\ee
where $\tanh\gamma = (Q_\text{L}-Q_\text{R})/(Q_\text{L}+Q_\text{R}).$ Repeating the derivation for $\vev{\Theta(-s)(-s)}$ and using Eqs. \eqf{Mt}, we arrive at Eq. \eqf{RES1} for $ M(x,t).$

\section{Alternative derivation}
\label{sec:alt}

In this appendix we provide an alternative derivation of $ M(x,t)$ yielding the expression \eqf{res}. The starting point is Eq. \eqf{Meqs} but now we compute the expectation values based on the probability that the number of net crossings $s$ admits a given value in a configuration.

A straight line from the left of momentum $p$ can intersect the $A=[(0,0),\,(x,t)]$  segment (see Fig. \ref{sketch}) only if its velocity is greater than $x/t.$ The probability that it intersects $A$ is, due to the even spatial distribution of lines, given by the length of the interval where the line can come from divided by the length $L_\text{L}$ of the left system, $|x-v_pt|/L.$ Similarly, a line from the right can cross $A$ if $v_p<x/t$ with probability  $(x-v_pt)/L_\text{R}.$ Then the probability that a randomly chosen straight line from the left or the right intersects the segment $A$ is
\begin{align}
q_\text{L} &= n_\text{L}^{-1}\int\frac{\ud p}{2\pi}\, \Theta(v_p-x/t)f_\text{L}(p) \, \frac{v_pt-x}{L_\text{L}}=Q_\text{L}/N_\text{L}\,,\\
q_\text{R} &= n_\text{R}^{-1}\int\frac{\ud p}{2\pi}\, \Theta(x/t-v_p)f_\text{R}(p) \, \frac{x-v_pt}{L_\text{R}}=Q_\text{R}/N_\text{R}\,,
\end{align}
where $N_\text{L/R}$ are the total initial particle numbers on the left and on the right, and $Q_\text{L/R}$ are defined in Eqs. \eqf{Qdef}.

The key quantity in the calculation is the net crossing number $s=k_\text{L}-k_\text{R},$ where $k_\text{L}$ and $k_\text{R}$  denote the number of left and right crossing lines in a configuration. The probability of such a configuration, thanks to the independence of the straight lines, is
\be
P(k_\text{L},k_\text{R}) = \binom{N_\text{L}}{k_\text{L}}\binom{N_\text{R}}{k_\text{R}}
q_\text{L}^{k_\text{L}}(1-q_\text{L})^{N_\text{L}-k_\text{L}}q_\text{R}^{k_\text{R}}(1-q_\text{R})^{N_\text{R}-k_\text{R}}\,.
\ee
Then $ M(x,t)$ in Eq. \eqf{Mt} can be computed as
\be
 M(x,t)=\sum_{k_\text{L}=0}^{N_\text{L}}\sum_{k_\text{R}=0}^{N_\text{R}}
P(k_\text{L},k_\text{R})\,
(k_\text{L}-k_\text{R})[\Theta(k_\text{L}-k_\text{R}){\hat\mu_\text{L}}+\Theta(k_\text{R}-k_\text{L}){\hat\mu_\text{R}}\big]\,.
\ee
Now we use the identity
\be
x[\Theta(x){\hat\mu_\text{R}}+\Theta(-x){\hat\mu_\text{L}}] = 
\frac12({\hat\mu_\text{R}}+{\hat\mu_\text{L}})x+\frac12({\hat\mu_\text{R}}-{\hat\mu_\text{L}})\,|x|\,,
\ee
and we rewrite $ M(x,t)$ as
\be
\begin{split}
 M(x,t)=&\frac12({\hat\mu_\text{R}}+{\hat\mu_\text{L}}) (N_\text{L}q_\text{L}-N_\text{R}q_\text{R})\\
+&\frac12({\hat\mu_\text{R}}-{\hat\mu_\text{L}})
\sum_{k_\text{L}=0}^{N_\text{L}}\sum_{k_\text{R}=0}^{N_\text{R}}\binom{N_\text{L}}{k_\text{L}}\binom{N_\text{R}}{k_\text{R}}
q_\text{L}^{k_\text{L}}(1-q_\text{L})^{N_\text{L}-k_\text{L}}q_\text{R}^{k_\text{R}}(1-q_\text{R})^{N_\text{R}-k_\text{R}}|k_\text{L}-k_\text{R}|\,.
\end{split}
\ee

Without the loss of generality we can assume that $N_\text{L}=N_\text{R}=N$ since this can be achieved by setting the ratio of the lengths of the two segments which however should not matter in the thermodynamic limit. It turns out that the double sum can be rewritten as
\be
S=N(q_\text{L}+q_\text{R})-N\sum_{l=0}^{N-1}\sum_{k=0}^{N-l-1}\binom{2l}{l}\frac2{l+1}\binom{k+2l}{2l}(q_\text{L}q_\text{R})^{l+1}(1-q_\text{L}-q_\text{R})^k\,,
\ee
which has the advantage that it depends on two combinations, $q_\text{L}q_\text{R}$ and $1-q_\text{L}-q_\text{R},$ moreover, one of the sums can be computed analytically:
\be
\sum_{k=0}^{N-l-1}\binom{k+2l}{2l}(1-q_\text{L}-q_\text{R})^k= \frac1{(q_\text{L}+q_\text{R})^{2l+1}}\left[1-(N-l)\binom{N+l}{2l}B_{1-q_\text{L}-q_\text{R}}(N-l,2l+1)\right]\,,
\ee
where $B_z(a,b)$ is the incomplete Euler beta function. 

In the thermodynamic limit,
\be
\lim_{N\to\infty}\sum_{k=0}^{N-l-1}\binom{k+2l}{2l}(1-Q_\text{L}/N-Q_\text{R}/N)^k 
= \left[1-\frac{\Gamma(2l+1,Q_\text{L}+Q_\text{R})}{(2l)!}\right]\left(\frac{N}{Q_\text{L}+Q_\text{R}}\right)^{2l+1}\,,
\ee
where $\Gamma(a,y)=\int_y^\infty\ud z z^{a-1}e^{-z}$ is the incomplete gamma function, so we obtain
\be
\lim_{N\to\infty} S = (Q_\text{R}+Q_\text{L})-(Q_\text{R}+Q_\text{L})\sum_{l=0}^\infty\binom{2l}{l}\frac2{l+1}\left(\frac{Q_\text{L}Q_\text{R}}{(Q_\text{L}+Q_\text{R})^2}\right)^{l+1} \left[1-\frac{\Gamma(2l+1,Q_\text{L}+Q_\text{R})}{(2l)!}\right]\,,
\ee
where we took $N$ to infinity in the upper limit of the sum as all the explicit dependence of the summand on $N$ has disappeared. It is convenient to introduce the notations
\be
r=Q_\text{R}+Q_\text{L}\,,\qquad p=\sqrt{Q_\text{L}Q_\text{R}}\,,\qquad R=\sqrt{Q_\text{R}}-\sqrt{Q_\text{L}}
\ee
where $R^2=r-2p.$  The first term in the bracket gives in the sum
\be
\sum_{l=0}^\infty\binom{2l}{l}\frac2{l+1}\left(\frac{p}r\right)^{2l+2} = 1-\sqrt{1-4(p/r)^2}\,,
\ee
while the for the second one we obtain by switching the sum and the integral in the definition of the Gamma function
\be
\sum_{l=0}^\infty\binom{2l}l\frac2{l+1}\left(\frac{p}r\right)^{2l+2}\frac1{(2l)!}\Gamma[2l+1,r]= 
\left(\frac{p}r\right)^2\int_\text{R}^\infty\ud z e^{-z} \sum_{l=0}^\infty\frac1{l!^2}\frac2{l+1}\left(\frac{p}rz\right)^{2l}
= 2\frac{p}r\int_\text{R}^\infty\ud z \frac{e^{-z}}z I_1(2p/r\cdot z)\,,
\ee
where $I_1(x)$ is the modified Bessel function of the first kind. So we arrive at
\be
\lim_{N\to\infty} S 
= s\sqrt{1-4(p/r)^2}+
2p\int_\text{R}^\infty\ud z\, \frac{e^{-z}}z I_1(2p/r\cdot z)
=|Q_\text{R}-Q_\text{L}|+2p
\int_1^\infty\ud z\, \frac{e^{-rz}}z \,I_1(2pz)\,.
\ee

Collecting the terms, we finally obtain in the thermodynamic limit
\begin{multline}
 M(x,t) = \frac12({\hat\mu_\text{R}}+{\hat\mu_\text{L}})(Q_\text{L}-Q_\text{R})
+\frac12({\hat\mu_\text{R}}-{\hat\mu_\text{L}})
\left[|Q_\text{R}-Q_\text{L}|+2p \int_1^\infty\ud z\, \frac{e^{-sz}}z \,I_1(2pz)\right] \\
=(Q_\text{R}-Q_\text{L})\left[\Theta(Q_\text{R}-Q_\text{L}){\hat\mu_\text{R}}+\Theta(Q_\text{L}-Q_\text{R}){\hat\mu_\text{L}}\right]+
({\hat\mu_\text{R}}-{\hat\mu_\text{L}})\sqrt{Q_\text{L}Q_\text{R}}
\int_1^\infty\ud z\, \frac{e^{-(Q_\text{R}+Q_\text{L})z}}z \,I_1\left(2\sqrt{Q_\text{L}Q_\text{R}}z\right)\,.
\end{multline}

\section{Some asymptotic expressions}
\label{app:asym}

In this appendix we list approximating expressions valid for large times and used to derive Eq. \eqf{jmas} of the main text. Using $Q_\text{R/L}\approx t$ we find
\bes
\label{besselas}
\begin{align}
e^{-(Q_\text{R}+Q_\text{L})}I_1(2\sqrt{Q_\text{R}Q_\text{L}})&\approx \frac{e^{-\left(\sqrt{Q_\text{R}}-\sqrt{Q_\text{L}}\right)^2}}{(Q_\text{L}Q_\text{R})^{1/4}\sqrt{4\pi}} \,,\\
\int_1^\infty\ud z\, \frac{e^{-(Q_\text{R}+Q_\text{L})z}}z\,I_1(2\sqrt{Q_\text{R}Q_\text{L}}z)&\approx
\frac{e^{-\left(\sqrt{Q_\text{R}}-\sqrt{Q_\text{L}}\right)^2}-\sqrt\pi \left|\sqrt{Q_\text{R}}-\sqrt{Q_\text{L}}\right|\,\mathrm{erfc}\left(\left|\sqrt{Q_\text{R}}-\sqrt{Q_\text{L}}\right|\right)}{(Q_\text{R}Q_\text{L})^{1/4}\sqrt{\pi}}\,,\\
\int_1^\infty\ud z\, e^{-(Q_\text{R}+Q_\text{L})z}\,I_1(2\sqrt{Q_\text{R}Q_\text{L}}z) &\approx 
\frac{\mathrm{erfc}\left(\left|\sqrt{Q_\text{R}}-\sqrt{Q_\text{L}}\right|\right)}{2 \left|\sqrt{Q_\text{R}}-\sqrt{Q_\text{L}}\right|(Q_\text{R}Q_\text{L})^{1/4}}\,.
\end{align}
\esu

\section{S-matrix of the $O(3)$ non-linear sigma model}
\label{app:Smatrix}

The S-matrix in the $x,y,z$ spin component basis is given by \cite{Zamolodchikov1979}
\be
S_{\alpha\beta}^{\gamma\delta} = 
\sigma_1(\theta)\delta_{\alpha\beta}\delta^{\gamma\delta} +
\sigma_2(\theta)\delta_\alpha^\gamma\delta_\beta^\delta+
\sigma_3(\theta)\delta_\alpha^\delta\delta_\beta^\gamma\,,
\ee
where
\begin{align}
\sigma_1(\th) = \frac{2i\pi\th}{(\th+i\pi)(\th-2i\pi)}\,,\\
\sigma_2(\th) = \frac{\th(\th-i\pi)}{(\th+i\pi)(\th-2i\pi)}\,,\\
\sigma_3(\th) = \frac{-2i\pi(\th-i\pi)}{(\th+i\pi)(\th-2i\pi)}\,.
\end{align}
This is the basis where the $SU(2)$ generators have the form
\be
J_1=i\begin{pmatrix}
0&0&0\\
0&0&-1\\
0&1&0
\end{pmatrix}\,,
\qquad
J_2=i\begin{pmatrix}
0&0&1\\
0&0&0\\
-1&0&0
\end{pmatrix}\,,
\qquad
J_3=i\begin{pmatrix}
0&-1&0\\
1&0&0\\
0&0&0
\end{pmatrix}\,.
\ee
A unitary transformation $I_j = UJ_jU^{-1}$ with
\be
U=i\begin{pmatrix}
-1/\sqrt{2}&i/\sqrt{2}&0\\
0&0&1\\
1/\sqrt{2}&i/\sqrt{2}&0
\end{pmatrix}
\ee
brings these to the form in the ``$m$-basis'':
\be
I_1=\frac1{\sqrt{2}}\begin{pmatrix}
0&1&0\\
1&0&1\\
0&1&0
\end{pmatrix}\,,
\qquad
I_2=\frac{-i}{\sqrt{2}}\begin{pmatrix}
0&1&0\\
-1&0&1\\
0&-1&0
\end{pmatrix}\,,
\qquad
I_3=\begin{pmatrix}
1&0&0\\
0&0&0\\
0&0&-1
\end{pmatrix}\,.
\ee
This means that the relation between the two bases
\begin{align}
\{|\alpha\rangle_{xyz} \} &= \{|1\rangle_{xyz}\,,|2\rangle_{xyz}\,,|3\rangle_{xyz}\} 
= \{|x\rangle\,,|y\rangle\,,|z\rangle\}\,,\\
\{|j\rangle_m \} &= \{|1\rangle_m\,,|2\rangle_m\,,|3\rangle_m\} 
= \{|+\rangle\,,|0\rangle\,,|-\rangle\}
\end{align}
 is given by
\be
|\alpha\rangle_{xyz} = U_{j\alpha}|j\rangle_m\,,\qquad |j\rangle_m = (U^{-1})_{\alpha j}|\alpha\rangle_{xyz}\,.
\ee
The S-matrix acts in the tensor product space so in the $m$-basis it is given by
\be
S_{ij}^{kl} = (U^{-1})_{\alpha i}(U^{-1})_{\beta j}U_{k\gamma}U_{l\delta}\;S_{\alpha\beta}^{\gamma\delta}\,.
\ee
This way we obtain
\begin{align}
S_{++}^{++} = S_{--}^{--} &= \sigma_2+\sigma_3 = \frac{\th-i\pi}{\th+i\pi}\,,\\
S_{+0}^{+0} = S_{0+}^{0+} = S_{-0}^{-0} = S_{0-}^{0-}& = \sigma_2 = \frac{\th(\th-i\pi)}{(\th+i\pi)(\th-2i\pi)}\,,\\
S_{+0}^{0+} = S_{0+}^{+0} = S_{-0}^{0-} = S_{0-}^{-0}& = \sigma_3 = \frac{-2i\pi(\th-i\pi)}{(\th+i\pi)(\th-2i\pi)}\,,\\
S_{+-}^{+-} = S_{-+}^{-+} & = \sigma_1 + \sigma_2 = \frac{\th}{\th-2i\pi} \,,\\
S_{+-}^{-+} = S_{-+}^{+-} & = \sigma_1 + \sigma_3 = \frac{-2\pi^2}{(\th+i\pi)(\th-2i\pi)}\,,\\
S_{+-}^{00} = S_{-+}^{00} = S_{00}^{+-} = S_{00}^{-+} & = -\sigma_1  = \frac{-2i\pi\th}{(\th+i\pi)(\th-2i\pi)}\,,\\
S_{00}^{00} &= \sigma_1 + \sigma_2 + \sigma_3 \,.
\end{align}
Note that according to the notation convention for the S-matrix, $S_{+0}^{0+}=\dots=\sigma_3,$ $S_{+-}^{-+}=S_{-+}^{+-}=\sigma_1+\sigma_3$ describe reflections, $S_{+0}^{+0}=\dots=\sigma_2,$ $S_{+-}^{+-}=S_{-+}^{-+}=\sigma_1+\sigma_2$ describe transmissions, and particle flavor changing scatterings in the neutral channel are described by $S_{+-}^{00}=\dots=-\sigma_1.$ The S-matrix satisfies the symmetry relations due to $P$, $C$, and $T$ invariance

\be
S_{ij}^{kl}(\th) = S_{ji}^{lk}(\th) = S_{\bar i\bar j}^{\bar k\bar l}(\th) = S_{lk}^{ji}(\th)\,,
\ee
the unitarity and crossing relations
\be
S_{ij}^{nm}(\th)S_{nm}^{kl}(-\th)=\delta_i^k\delta_j^l\,,\qquad S_{ij}^{kl}(\th) = S_{i\bar l}^{k\bar j}(i\pi-\th)
\ee
as well as the Yang--Baxter equation
\be
S_{ij}^{\beta\alpha}(\th_{12})S_{\beta k}^{n\gamma}(\th_{13})S_{\alpha\gamma}^{ml}(\th_{23})=
S_{jk}^{\beta\gamma}(\th_{23})S_{i\gamma}^{\alpha l}(\th_{13})S_{\alpha\beta}^{nm}(\th_{12})\,.
\ee
In matrix notation,
\bigskip
\be
S = 
\left(
\begin{tabular}{c|cc|ccc|cc|c}
$\sigma_2+\sigma_3$&0&0&0&0&0&0&0&0\\
\hline
0&$\sigma_2$&$\sigma_3$&0&0&0&0&0&0\\
0&$\sigma_3$&$\sigma_2$&0&0&0&0&0&0\\
\hline
0&0&0&$\sigma_1+\sigma_2$&$-\sigma_1$&$\sigma_1+\sigma_3$&0&0&0\\
0&0&0&$-\sigma_1$&$\sigma_1+\sigma_2+\sigma_3$&$-\sigma_1$&0&0&0\\
0&0&0&$\sigma_1+\sigma_3$&$-\sigma_1$&$\sigma_1+\sigma_2$&0&0&0\\
\hline
0&0&0&0&0&0&$\sigma_2$&$\sigma_3$&0\\
0&0&0&0&0&0&$\sigma_3$&$\sigma_2$&0\\
\hline
0&0&0&0&0&0&0&0&$\sigma_2+\sigma_3$
\end{tabular}
\right)
\ee
where the 2-particle basis is
\begin{center}
\begin{tabular}{ccccccccc}
$|++\rangle;$&$|+0\rangle,$&$|0+\rangle;$&$|+-\rangle,$&$|00\rangle,$&$|-+\rangle;$&$|-0\rangle,$&$|0-\rangle;$&$|--\rangle.$
\end{tabular}
\end{center}
As $\th\to0,$
\be
\sigma_1\to0\,,\qquad \sigma_2 \to 0\,, \qquad \sigma_3 \to -1\,,
\ee
so all scatterings become purely reflective with transmissions and spin changing scatterings suppressed.

\section{Details on the numerical simulations}

In this section we discuss in more detail the numerical algorithm used. It consists of mainly two distinct steps: $(i)$ Generation of many semiclassical configurations. In each such  configuration we keep track of the space-time trajectories for the quasiparticles. $(ii)$ Statistical averages over many configurations to determine the evolution in time of the spatial profiles of various quantities of interest such as the magnetization. 

\emph{(i) Generation of a single semiclassical configuration}: A semiclassical configuration (a typical one is displayed in Fig.~\ref{sketch}) consists of the space-time trajectories and initial spins of the particles.  To generate one,  we first divide our physical system into two subsystems of equal size $L/2$ labeled as L and R. The coordinate $x$ and the length of the system is measured in units of Compton length, $l_d = \hbar c /\Delta$, while the time coordinate is measured in units of $t_0 = \hbar/\Delta$. In our numerics, the typical system sizes are $L\simeq 10^4-10^5$, which guarantees that for times $t\apprle 10^3$ only a small fraction ($\lesssim 2 \%$) of the total number of particles escapes at the boundaries. At any moment in time $t$, each particle $j$ is characterized  by a coordinate $x_j(t),$ a momentum $p_j(t),$ and a spin variable $\mu_j(t).$  The momenta and the spins are initially drawn from the distribution
 \be
\label{fmuptherm1}
P_{\alpha}(\mu,p) \sim  e^{-\beta_{\alpha}\eps(p)} g(\mu) ,\qquad\qquad \alpha = \{\text{L, R}\}\,,
\ee
where the discrete normalized probability distribution $g(\mu)$ is fixed by the average spin. Note that neither the spin nor the momentum distribution needs to be thermal. For the sake of simplicity, however, we have assumed a thermal momentum distribution, and considered non-thermal distributions only in the spin variables in our numerical simulations.

Notice that~\eqref{fmuptherm1} factorizes in the momentum and spin components so the two variables are initially independent. As the temperatures $\beta^{-1}_\text{L/R}$ of the two subsystems are different,
the initial particle densities on the two sides are also different. The particle densitites are evaluated according to  Eq.~\eqref{eq:n_init}. Once concentrations are fixed, we generate randomly their positions and index them from $1$ to $N_\text{L/R}.$
In this way the semiclassical configuration at $t=0$ is fully constructed  as each  particle is fully characterized by its coordinate $x_j(0)$, momentum $p_j(0)$ and 
spin $\mu_j(0)$ with $1\le j\le N_\text{L/R}$. At later times, $t>0$, particles moves with constant velocities in between the collisions and their trajectories are described as rays in the $(x,t)$ 
plane, as depicted in Fig.~\ref{sketch}. A crossing of two rays signals a collision of two 
particles. By simple geometrical arguments we  determine all intersection coordinates
$\{x_I, t_I\}$ and order them chronologically. We also keep track of the labels of the particles 
that enter the collision. When two such particles collide, they exchange their momenta as the masses of all particles are equal.  
Furthermore, in the universal limit, characterized by the fully reflective S-matrix, their spins remain unaltered too. Numerically, we are able to go beyond the universal limit and allow for transmission in the spin sector. In this non-universal limit, transmission/reflection probability at each collison is encoded in the components of the S-matrix, as discussed in Appendix~\ref{app:Smatrix}. In this work, instead of determining the full spin wave function \cite{Moca2016}, we use a simple Monte Carlo sampling to decide the outcome of each collision event with a given probability. 
 Consequently, starting with an initial configuration $\{x_j(0), p_j(0), \mu_j(0)\}, j=1\dots N_\text{L/R}$, we can determine at any later time $t>0$ the full configuration 
$\{x_j(t), p_j(t), \mu_j(t)\}, j=1\dots N_\text{L/R}$, of all particles in terms of their position, momenta and spins. 

\emph{(ii) Statistical averages:} To represent the magnetization profiles or the time evolution of the relative density at the interface, we first collect data by sampling ~$10^3-10^4$ configurations, and then perform a statistical analyis to measure the quantity of interest. For example, if we want to measure the magnetization profile at a given time $t_s$ (see for example  the magnetization profile in  Fig.~\ref{fig:mprofiles}) we 
determine both the positions and the spins of all the particles for each configuration at time $t_s$.  Then we perform an average of the magnetization over all the configurations using histograms.

\end{document}